\documentclass[useAMS,usenatbib]{mn2e}

\def\sn{SN~2008iy}
\def\kms{km~s$^{-1}$}
\def\swift{\emph{Swift}}
\def\arcmin{\hbox{$^\prime$}}
\def\arcsec{\hbox{$^{\prime\prime}$}}
\def\msun{M$_\odot$}
\def\lsun{L$_\odot$}

\def\jhk{{\em J}, {\em H}, and {\em K$_s$}}

\def\mdot{$\dot{M}$}
\newcommand{\bvri}{\protect\hbox{$BV\!RI$}}

\newcommand {\aap}{A\&A}
\newcommand {\apj}{ApJ}
\newcommand {\apjl}{ApJL}
\newcommand {\apjs}{ApJS}
\newcommand {\aj}{AJ}
\newcommand {\nat}{Nat}
\newcommand {\mnras}{MNRAS}
\newcommand {\pasp}{PASP}
\newcommand {\iaucirc}{IAU Circ.}
\newcommand {\aaps}{A\&AS}
\newcommand {\araa}{ARA\&A}

\usepackage{natbib,epsfig,graphicx,rotating,lscape,rotating}

\title[The 400 Day Rise of \sn]{SN~2008iy: An Unusual Type IIn Supernova 
with an Enduring 400 Day Rise Time}

\author[Miller et al.]{A. A. Miller$^{1}$\thanks{E-mail: amiller@astro.berkeley.edu.}, 
J. M. Silverman$^{1}$,
N. R. Butler$^{1}$,
J. S. Bloom$^{1}$\thanks{Sloan Research Fellow.}, 
R. Chornock$^{1,2}$,   
\newauthor A. V. Filippenko$^{1}$,
M. Ganeshalingam$^{1}$,
C. R. Klein$^{1}$,
W. Li$^{1}$, 
P. E. Nugent$^{3}$,
\newauthor N. Smith$^{1}$, 
and T. N. Steele$^{1}$. \\
$^1$Department of Astronomy, University of California, Berkeley, CA 94720-3411, USA. \\
$^2$Harvard-Smithsonian Center for Astrophysics, 60 Garden Street, Cambridge, MA 02138, USA. \\
$^3$Lawrence Berkeley National Laboratory, 1 Cyclotron Road, Berkeley, CA 94720, USA.
}

\begin{document}

\date{}

\pagerange{\pageref{firstpage}--\pageref{lastpage}} \pubyear{2009}

\maketitle

\label{firstpage}

\begin{abstract}

We present spectroscopic and photometric observations of the Type IIn 
supernova (SN) 2008iy. \sn\ showed an unprecedentedly long rise time of 
$\sim$400 days, making it the first known SN to take significantly longer than 
100 days to reach peak optical luminosity. The peak absolute magnitude 
of \sn\ was $M_r \approx -19.1$ mag, and the total radiated 
energy over the first $\sim$700 days was $\sim$2 $\times 10^{50}$ erg. 
Spectroscopically, \sn\ is 
very similar to the Type IIn SN~1988Z at late times, and, like SN~1988Z, 
it is a luminous X-ray source (both supernovae had an X-ray luminosity 
$L_{\rm X} > 10^{41}$ erg s$^{-1}$). \sn\ has a growing near-infrared excess 
at late times similar to several other SNe~IIn. The H$\alpha$ emission-line profile of 
\sn\ shows a narrow P Cygni absorption component, 
implying a pre-SN wind speed of $\sim$100 \kms. We argue 
that the luminosity of \sn\ is powered via the interaction of the SN ejecta 
with a dense, clumpy circumstellar medium. The $\sim$400 day rise 
time can be understood if the number density of clumps increases with
distance over a 
radius $\sim$1.7 $\times 10^{16}$ cm from the progenitor. This scenario is 
possible if the progenitor experienced an episodic phase of enhanced 
mass loss $<$ 1 century prior to explosion or if the progenitor wind speed 
increased during the decades before core collapse. We favour the former 
scenario, which is reminiscent of the eruptive mass-loss episodes 
observed for luminous blue variable (LBV) stars. The progenitor wind 
speed and increased mass-loss rates serve as further evidence that at least 
some, and perhaps all, Type IIn supernovae experience LBV-like eruptions 
shortly before core collapse. We also discuss the host galaxy of \sn, a 
subluminous dwarf galaxy, and offer a few reasons why the recent 
suggestion that unusual, luminous supernovae preferentially occur in dwarf 
galaxies may be the result of observational biases.

\end{abstract}


\section{Introduction}

The recent development of synoptic, wide-field imaging has revealed an 
unexpected diversity of transient phenomena. One such example is the 
discovery of a new subclass of very luminous supernovae (VLSNe; e.g.,  
\citealt{ofek06gy,smith07-2006gy,quimby05ap}). While events of this nature 
are rare \citep{miller08es,quimby2009}, each new discovery serves to bracket 
our understanding of the physical origin of well-established classes of 
SNe and does, in principle, demand an increased clarity in our 
understanding of the post-main-sequence evolution of massive stars.

Yet another unusual transient was discovered by the 
Catalina Real-Time Transient Survey (CRTS; \citealt{drake2009}), which 
they announced via an ATel as CSS080928:160837+041626 on 2008 Oct.\ 07 
UT\footnote{UT dates are used throughout this paper unless 
otherwise noted.} \citep{drake08-2008iy}. The transient was classified as a 
Type IIn SN with a spectrum taken on 2009 Mar.\ 27 (\citealt{mahabal2009}; see 
\citealt{schlegel96} for a definition of the SN~IIn subclass, and 
\citealt{filippenko1997} for a review of the spectral properties of SNe). 
The SN was later given 
the IAU designation \sn\ \citep{catelan2009}. \citet{mahabal2009} noted 
that the transient was present on CRTS images dating back to 2007 Sep.\ 13;
however, it went undetected by the CRTS automated transient detection 
software until 2008 because it was blended with a non-saturated, 
nearby ($\sim$11\arcsec\ separation) star. 

Here, we present our observations and analysis of \sn, which peaked 
around 2008 Oct.\ 29 
\citep{catelan2009} and had a rise time of $\sim$400 days. This implies that 
\sn\ took 
longer to reach peak optical brightness than any other known SN. Type II SN 
rise times are typically $\la$ 1 week (e.g., SNe~2004et and 2006bp, 
\citealt{wli-2004et, quimby06bp}; see also \citealt{patat1993,wli-rates-LF}), 
and have never previously been 
observed to rise $\ga$ 100 days, let alone 400, making \sn\ another 
rare example of the possible outcomes for the end of the stellar life cycle. 
In addition to an extreme rise time, \sn\ is of great interest because 
the unique circumstellar medium (CSM) in which it exploded may 
provide a link to very long-lived SNe, such as SN 1988Z, and thus 
provide clues into the nature of their progenitors. 

This paper is organised as follows. Section 2 presents the observations.
The data are analysed in Section 3, and the results are
discussed in Section 4. We give our conclusions, as well as 
predictions for the future behaviour of \sn, in Section 5. 

\section{Observations}

\subsection{Photometry}

\begin{figure*}
\begin{center}
\includegraphics[width=160mm]{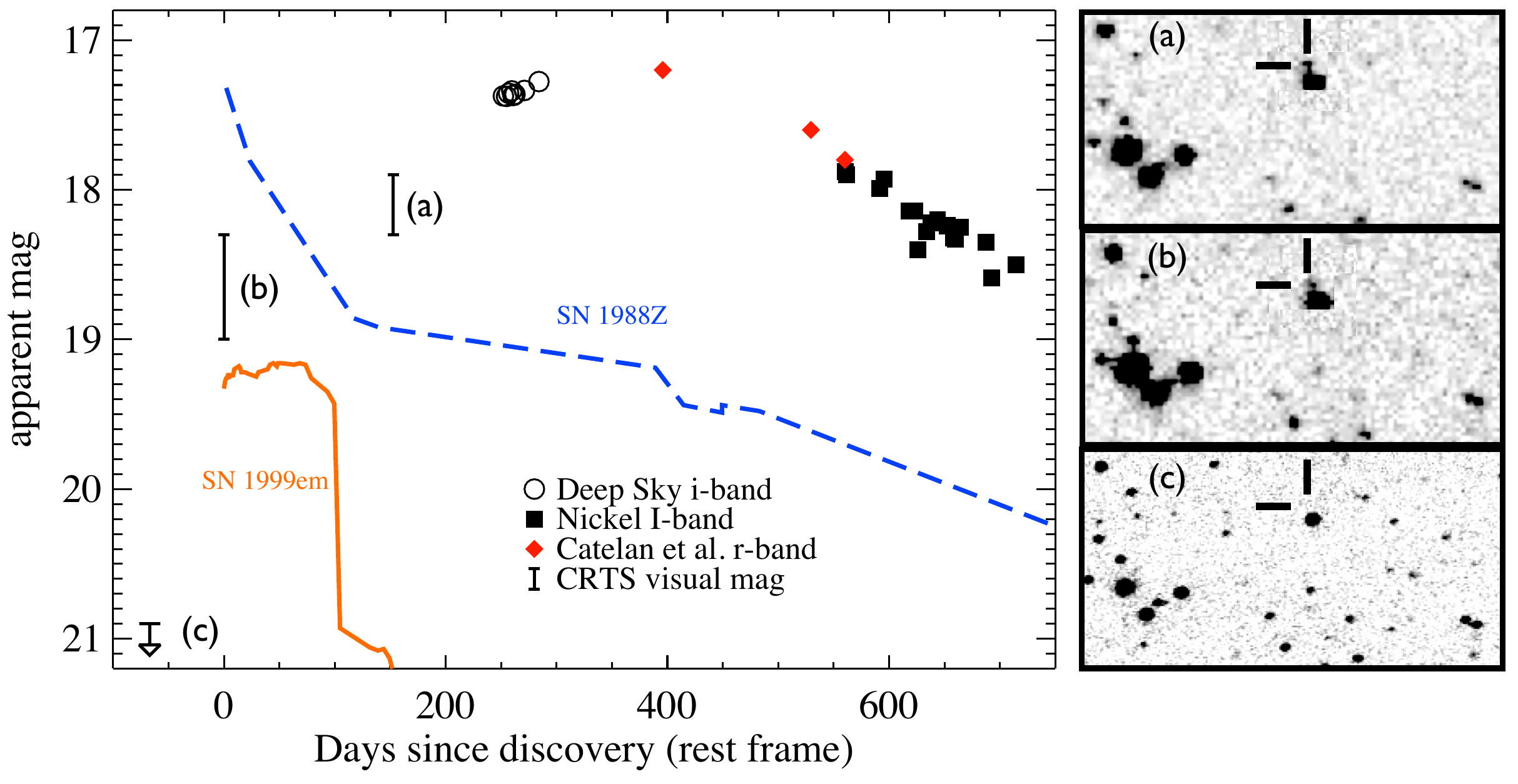}
\caption{{\it Left}: Apparent optical light curve of \sn\ showing the long rise time of 
the SN, including data from Deep Sky 
(this work), the Nickel telescope (this work), \citet{catelan2009}, and the 
CRTS website. Note: conservatively, we take the start time of the SN 
($t =0$) to coincide with the first epoch where CRTS detects the source; 
however, the true explosion date is likely prior to 2007 Sep.\ 13, between 
the last Deep Sky non-detection and the first CRTS detection. Notice that 
the light curve is very broad, and almost symmetric about the peak. For 
comparison we show the light curve of the standard Type II-P SN~1999em 
(data from \citealt{leonard02-99em}) and the Type IIn SN~1988Z, which shows 
many similarities to \sn\ at late times (see Section 3.2; SN~1988Z data from 
\citealt{turatto1993}), as they would have appeared at the 
redshift of \sn. {\it Right}:  Three panels with images of the field of 
\sn. Each image is $\sim 2.5\arcmin 
\times 4\arcmin$, with north up and east to the left. Black crosshairs mark 
the location of the SN. The bottom image, marked (c), shows the last 
non-detection from the Deep Sky data. The middle (b) and top (a) images 
show the CRTS detections on day 0 and day 153, respectively;
\sn\ can clearly be seen in each of them.  The SN was not flagged 
as a transient in either the middle or top image by the CRTS software. 
} 
\label{long-lc}
\end{center}
\end{figure*}

The field of \sn, which is located at $\alpha$ = 16$^h$08$^m$37.27$^s$, 
$\delta$ = +04$^\circ$16$'$26.5$''$ (J2000), was imaged multiple 
times by the Palomar Quest survey, 
and those data have been reprocessed as part of the Deep Sky 
project\footnote{{http://supernova.lbl.gov/$\sim$nugent/deepsky.html.}} 
(DS; \citealt{nugent2009}). The best constraints on the explosion date 
come from DS imaging: in a coadd of two images from 2007 Jul.\ 05, we do not 
detect the SN down to a 3$\sigma$ limit of $i > 20.9$ mag. DS images are 
best approximated by the Sloan Digital Sky Survey (SDSS; \citealt{adelman08}) 
$i$-band filter. 
The field of \sn\ has well-calibrated SDSS photometry which we use 
to calibrate the DS images. The observed DS magnitudes are shown
in Fig.~\ref{long-lc} and summarised in Table~\ref{tab-ds}.

\begin{table}
\begin{minipage}{80mm}
  \centering
\caption{Deep Sky Observations of \sn. 
}
\begin{tabular}{rccc}
  \hline
  date & mag\footnote{Observed value; not corrected for Galactic extinction.} & $\sigma_{\rm mag}$\footnote{The calibration uncertainty 
dominates over the statistical uncertainty with values of $\sim$0.12 and 
$\sim$0.01 mag, respectively.} \\ 
  (MJD) &  & \\
  \hline
54618.35 & 17.37 &  0.12 \\
54621.35 & 17.38 &  0.12 \\
54623.33 & 17.35 &  0.12 \\
54626.32 & 17.34 &  0.12 \\
54627.36 & 17.37 &  0.12 \\
54628.29 & 17.37 &  0.12 \\
54629.30 & 17.36 &  0.12 \\
54638.21 & 17.34 &  0.12 \\
54651.20 & 17.28 &  0.12 \\
\hline
\end{tabular}
\label{tab-ds}
\end{minipage}
\end{table}

To constrain the rise time of \sn\ we visually estimate the 
possible range of magnitudes for the SN from images on the CRTS 
website\footnote{{http://crts.caltech.edu/.}} 
based on a comparison to SDSS images. As previously mentioned, \sn\ is 
blended with a nearby star and so the automated photometry produced by the 
CRTS does not detect the SN on these epochs. On 2007 Sep.\ 13 the SN was 
between 18.3 and 19.0 mag, while on 2008 Feb.\ 19 the SN was observed 
between 17.9 and 18.3 mag.

\begin{figure}
\includegraphics[width=84mm]{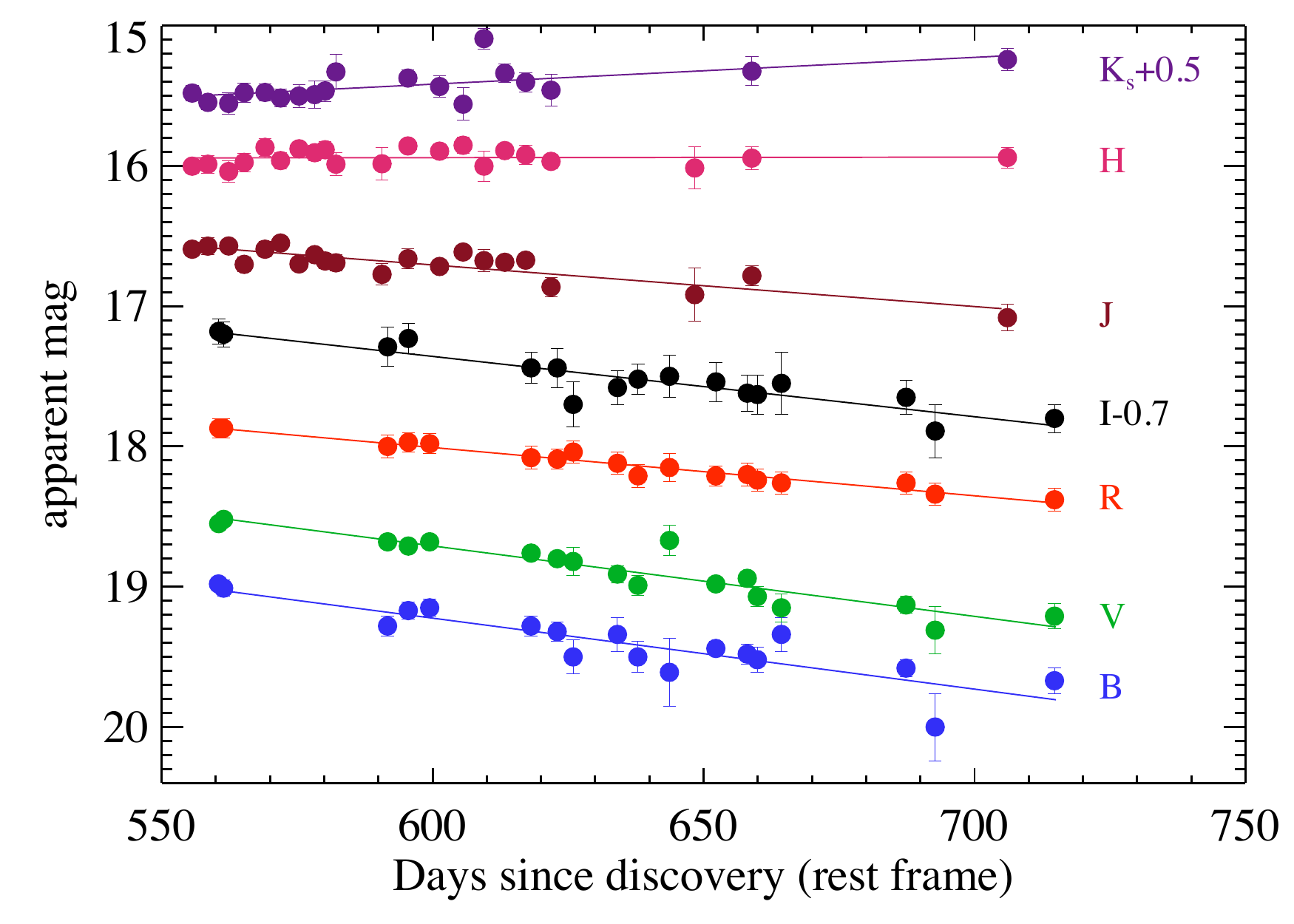}
\caption{Filtered photometry including \bvri\ from the Nickel telescope and 
\jhk\ from PAIRITEL, showing the decline rate of \sn\ between days 
$\sim$560 and 715. To help guide the eye, we illustrate the linear fit to each band 
used to determine the photometric decline rates (see text). Note that the 
single epoch of UVOT $B$ and $V$-band observations we measured, which 
agree to $< 1\sigma$ with Nickel data taken $<$ 2 days later, is not 
shown in this figure. 
} 
\label{filter-phot}
\end{figure}

Near-infrared (NIR) observations of \sn\ were conducted with the 1.3-m Peters 
Automated Infrared Imaging Telescope (PAIRITEL; \citealt{bloom06}) starting 
on 2009 Apr.\ 13. PAIRITEL observes simultaneously in the \jhk\ bands. 
Observations were scheduled and executed via a robotic 
system, and the data were reduced by an automated pipeline 
\citep{bloom06}. The SN flux was measured via aperture photometry
using SExtractor \citep{bertin}, calibrated 
against the Two Micron All Sky Survey \citep[2MASS][]{skrutskie-2mass}. Filtered 
photometry of \sn\ is shown in Fig.~\ref{filter-phot} and summarised in 
Table~\ref{tab-ptel}.

\begin{table}
\begin{minipage}{80mm}
  \centering
\caption{PAIRITEL Observations of \sn. 
}
\begin{tabular}{rccc}
\hline
$t_{\rm mid}$\footnote{Midpoint between the first and last exposures in a single stacked image.} & $J$ mag\footnote{Observed value; not corrected for Galactic extinction.} & $H$ mag{\Large $^b$} & $K_s$ mag{\Large $^b$} \\ 
(MJD) & (Vega) & (Vega) & (Vega) \\
\hline
54934.42 & 16.59 $\pm$  0.05 & 16.00 $\pm$  0.05 & 14.98 $\pm$  0.05 \\
54937.43 & 16.57 $\pm$  0.06 & 15.99 $\pm$  0.06 & 15.05 $\pm$  0.05 \\
54941.43 & 16.57 $\pm$  0.05 & 16.04 $\pm$  0.08 & 15.06 $\pm$  0.08 \\
54944.40 & 16.70 $\pm$  0.05 & 15.98 $\pm$  0.07 & 14.98 $\pm$  0.07 \\
54948.40 & 16.59 $\pm$  0.05 & 15.87 $\pm$  0.06 & 14.98 $\pm$  0.06 \\
54951.39 & 16.55 $\pm$  0.04 & 15.96 $\pm$  0.06 & 15.02 $\pm$  0.06 \\
54954.35 & 16.70 $\pm$  0.04 & 15.88 $\pm$  0.05 & 15.00 $\pm$  0.08 \\
54957.37 & 16.63 $\pm$  0.05 & 15.91 $\pm$  0.05 & 14.99 $\pm$  0.10 \\
54959.37 & 16.68 $\pm$  0.05 & 15.89 $\pm$  0.06 & 14.97 $\pm$  0.07 \\
54961.44 & 16.69 $\pm$  0.06 & 15.99 $\pm$  0.08 & 14.83 $\pm$  0.12 \\
54970.30 & 16.77 $\pm$  0.08 & 15.98 $\pm$  0.12 &  ...  \\
54975.28 & 16.66 $\pm$  0.07 & 15.86 $\pm$  0.05 & 14.87 $\pm$  0.06 \\
54981.35 & 16.72 $\pm$  0.04 & 15.90 $\pm$  0.05 & 14.93 $\pm$  0.08 \\
54986.29 & 16.61 $\pm$  0.05 & 15.85 $\pm$  0.06 & 15.06 $\pm$  0.12 \\
54990.30 & 16.68 $\pm$  0.08 & 16.00 $\pm$  0.11 & 14.59 $\pm$  0.07 \\
54994.30 & 16.69 $\pm$  0.05 & 15.89 $\pm$  0.04 & 14.84 $\pm$  0.06 \\
54998.31 & 16.67 $\pm$  0.05 & 15.92 $\pm$  0.07 & 14.90 $\pm$  0.07 \\
55003.18 & 16.86 $\pm$  0.07 & 15.97 $\pm$  0.05 & 14.96 $\pm$  0.11 \\
55030.20 & 16.92 $\pm$  0.19 & 16.01 $\pm$  0.15 &  ...  \\
55041.18 & 16.78 $\pm$  0.07 & 15.94 $\pm$  0.08 & 14.82 $\pm$  0.10 \\
55091.10 & 17.08 $\pm$  0.09 & 15.94 $\pm$  0.07 & 14.74 $\pm$  0.08 \\
\hline
\end{tabular}
\label{tab-ptel}
\end{minipage}
\end{table}

Ground-based optical observations of \sn\ were obtained using the 1.0-m 
Nickel telescope located at Lick Observatory (Mt. Hamilton, CA, USA) 
starting on 2009 Apr.\ 18. \bvri\ 
photometry from the Nickel was measured using the DAOphot package 
\citep{stetson1987} in 
IRAF\footnote{IRAF is distributed by the National Optical Astronomy 
Observatory, which is operated by the Association of Universities for 
Research in Astronomy (AURA) under cooperative agreement with the 
National Science Foundation.}, and transformed into the Johnson--Cousins 
system. Calibrations for the field were obtained on 
ten photometric nights with the Nickel telescope. The Nickel photometry 
is shown in Fig.~\ref{filter-phot} and summarised in 
Table~\ref{tab-nickel}.

\begin{table}
\begin{minipage}{80mm}
\caption{Nickel Observations of \sn. 
}
\begin{tabular}{rcccc}
\hline
date & $B$ mag\footnote{Observed value; not corrected for Galactic extinction.} & $V$ mag{\Large $^a$} & $R$ mag{\Large $^a$} & $I$ mag{\Large $^a$} \\
(MJD) & (Vega) & (Vega) & (Vega) & (Vega) \\
\hline
54939.52 & 19.06$\pm$ 0.01 & 18.59$\pm$ 0.03 & 17.86$\pm$ 0.01 & 17.87$\pm$ 0.03 \\ 
54940.46 & 19.09$\pm$ 0.03 & 18.56$\pm$ 0.02 & 17.86$\pm$ 0.02 & 17.89$\pm$ 0.03 \\ 
54971.37 & 19.35$\pm$ 0.06 & 18.71$\pm$ 0.04 & 17.99$\pm$ 0.04 & 17.98$\pm$ 0.11 \\ 
54975.36 & 19.25$\pm$ 0.03 & 18.75$\pm$ 0.02 & 17.96$\pm$ 0.02 & 17.92$\pm$ 0.07 \\ 
54979.46 & 19.23$\pm$ 0.03 & 18.71$\pm$ 0.02 & 17.97$\pm$ 0.02 &  ...  \\ 
54999.35 & 19.36$\pm$ 0.05 & 18.79$\pm$ 0.03 & 18.07$\pm$ 0.03 & 18.12$\pm$ 0.08 \\ 
55004.37 & 19.40$\pm$ 0.05 & 18.84$\pm$ 0.03 & 18.07$\pm$ 0.02 & 18.13$\pm$ 0.11 \\ 
55007.43 & 19.56$\pm$ 0.11 & 18.84$\pm$ 0.10 & 18.02$\pm$ 0.04 & 18.37$\pm$ 0.13 \\ 
55015.36 & 19.43$\pm$ 0.11 & 18.95$\pm$ 0.06 & 18.10$\pm$ 0.04 & 18.26$\pm$ 0.08 \\ 
55019.30 & 19.58$\pm$ 0.09 & 19.02$\pm$ 0.06 & 18.19$\pm$ 0.03 & 18.21$\pm$ 0.07 \\ 
55025.36 & 19.66$\pm$ 0.24 & 18.66$\pm$ 0.10 & 18.16$\pm$ 0.08 & 18.18$\pm$ 0.12 \\ 
55034.26 & 19.52$\pm$ 0.02 & 19.02$\pm$ 0.03 & 18.19$\pm$ 0.02 & 18.23$\pm$ 0.11 \\ 
55040.30 & 19.56$\pm$ 0.05 & 18.97$\pm$ 0.03 & 18.18$\pm$ 0.03 & 18.30$\pm$ 0.09 \\ 
55042.24 & 19.60$\pm$ 0.07 & 19.11$\pm$ 0.06 & 18.21$\pm$ 0.03 & 18.31$\pm$ 0.11 \\ 
55047.27 & 19.43$\pm$ 0.11 & 19.21$\pm$ 0.10 & 18.22$\pm$ 0.04 & 18.24$\pm$ 0.20 \\ 
55071.22 & 19.66$\pm$ 0.04 & 19.17$\pm$ 0.05 & 18.23$\pm$ 0.03 & 18.34$\pm$ 0.09 \\ 
55077.23 & 20.07$\pm$ 0.24 & 19.33$\pm$ 0.17 & 18.29$\pm$ 0.03 & 18.58$\pm$ 0.17 \\ 
55100.15 & 19.75$\pm$ 0.07 & 19.25$\pm$ 0.08 & 18.35$\pm$ 0.03 & 18.49$\pm$ 0.06 \\ 
\hline
\end{tabular}
\label{tab-nickel}
\end{minipage}
\end{table}

\sn\ was observed by the space-based \swift\ observatory on 2009 Apr.\ 15 
and 16. \swift\ observed the SN simultaneously with both the X-ray Telescope 
(XRT; \citealt{xrtref}) and the Ultraviolet/Optical Telescope 
(UVOT; \citealt{uvotref}). We downloaded the Level-2 UVOT images, and measured 
the $U$, $B$, and $V$ photometry using the recipe of \citet{li06}. The UV 
filters ($UVM1$, $UVW1$, and $UVW2$) were calibrated using the method of 
\citet{poole2008}. Our final UVOT photometry is summarised in 
Table~\ref{tab-uvot}. The XRT observed the field for a total of 10.71 ks. 
We extract the 0.3--10.0 keV counts from an extraction region of 64 pixels 
($\sim 2.5'$), where we fit the point-spread function model (see 
\citealt{butler2007}) at the centroid of the X-ray emission. From the native 
XRT astrometry the centroid of the X-ray emission, located at  
$\alpha$ = 16$^h$08$^m$37.23$^s$, $\delta$ = +04$^\circ$16$'$26.7$''$ (J2000), 
with a radial uncertainty of 5$''$ (90\% confidence interval), 
coincides with the optical position of \sn, suggesting that the X-rays 
are from the SN. In an astrometric fit relative to 2MASS we measure the 
position of the star $\sim$11\arcsec\ from the SN to be 
$\alpha$ = 16$^h$08$^m$36.98$^s$, $\delta$ = +04$^\circ$16$'$16.4$''$ (J2000), 
with a root-mean square scatter of 0.19\arcsec\ in $\alpha$ and 
0.18\arcsec\ in $\delta$ for our 
astrometric solution. This star is therefore well outside the 90\% confidence 
interval for the location of the X-ray emission. Furthermore, there 
is no catalogued X-ray source at the position of the SN in the ROSAT 
catalogue \citep{voges1999}, also 
suggesting that the star located $\sim$11\arcsec\ from the SN is not the 
source of X-ray emission. We measure 
a background-subtracted 0.3$-$10.0 keV count rate of (19 $\pm$ 6) 
$\times 10^{-4}$ counts s$^{-1}$. 

\begin{table}
\begin{minipage}{80mm}
  \centering
\caption{UVOT Observations of \sn. 
}
\begin{tabular}{rccc}
\hline
$t_{\rm mid}$\footnote{Midpoint between the first and last exposures in a single stacked image.} & filter & mag\footnote{Observed value; not corrected for Galactic extinction.} & $\sigma_{\rm mag}$ \\ 
(MJD) & & (Vega) & \\
\hline
54936.31 & UVW2 & 19.14 & 0.06 \\
54936.32 & UVM2 & 18.95 & 0.08 \\ 
54936.32 & UVW1 & 18.53 & 0.07 \\ 
54936.31 & U & 18.60 & 0.06 \\ 
54937.49 & UVW2 & 19.12 & 0.04 \\
54937.85 & B & 19.04 & 0.06 \\ 
54937.85 & V & 18.65 & 0.08 \\ 
\hline
\end{tabular}
\label{tab-uvot}
\end{minipage}
\end{table}

\subsection{Spectroscopy}

\begin{figure*}
\includegraphics[width=170mm]{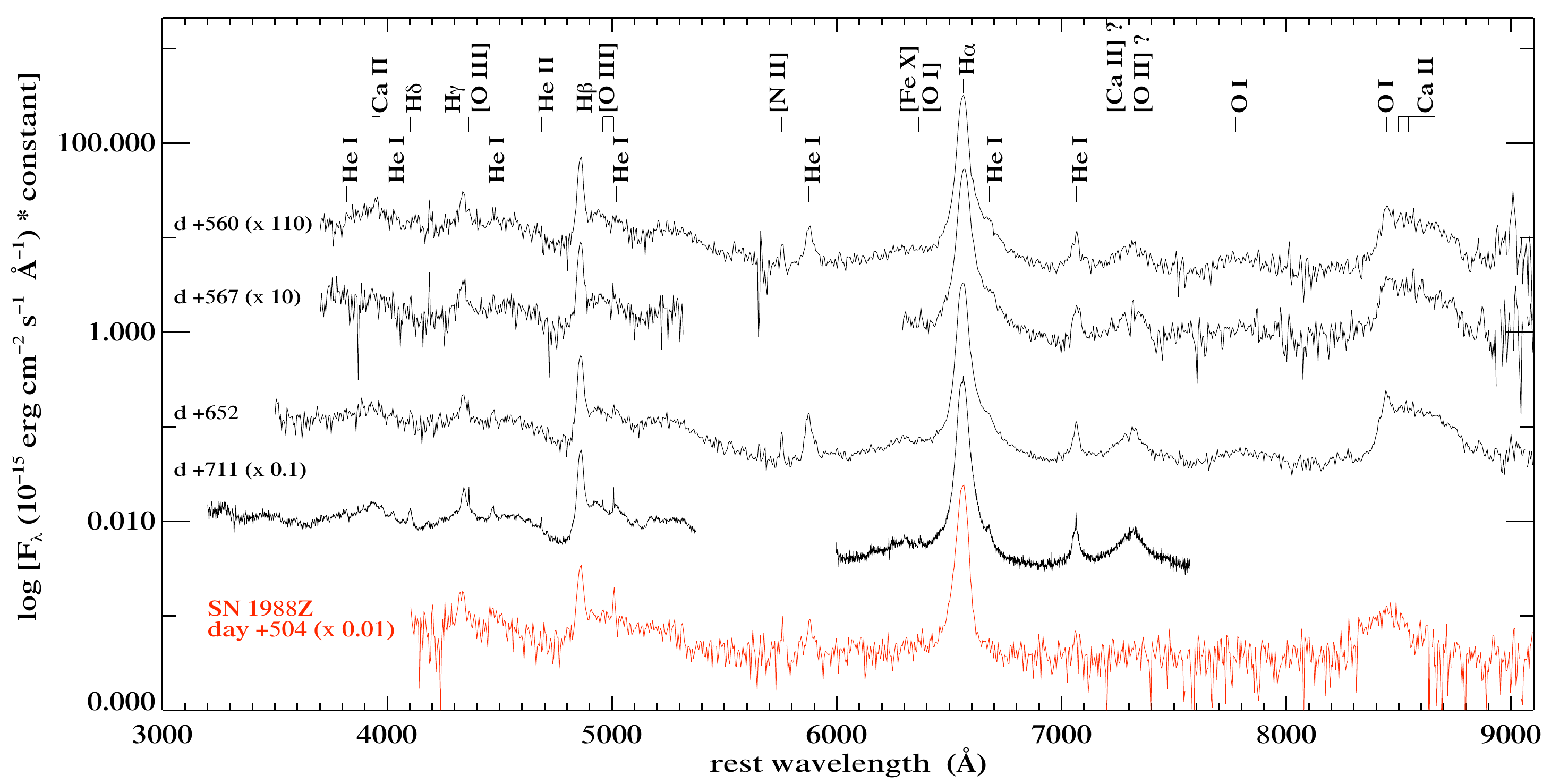}
\caption{Spectral sequence of \sn\ (shown in black) and a single, 
late-time spectrum of SN~1988Z from our spectral database (shown in red), 
which was observed 504 days after discovery. 
To the left of the spectra we give the time since discovery, in rest-frame 
days, at which each spectrum was taken. Prominent spectral features 
of \sn\ are labelled at the top of the figure. The spectra show little 
evolution between days 560 and 711. Fe~II multiplets 42, 48, and 49 
can also be seen, though we have not labelled those lines. The spectra 
have been corrected for Galactic reddening ($E(B-V) = 0.065$ mag; 
\citealt{sfd98}). We have assumed no reddening in the SN host galaxy (see 
Section 2.3).
} 
\label{spectra}
\end{figure*}

Spectra of \sn\ were obtained on 2009 Apr.\ 18, 2009 Apr.\ 25, and 
2009 Jul.\ 23 with the Kast spectrograph on the Lick 3-m Shane telescope 
\citep{kastref}. An additional spectrum was obtained on 2009 Sep.\ 22 
with the Low-Resolution Imaging Spectrometer (LRIS; \citealt{oke95}) on 
the 10-m Keck I telescope. To improve the spectral resolution at the 
location of the H$\alpha$ emission feature from \sn, we used the 1200 
lines mm$^{-1}$ grating on the red side of the double-arm (blue+red) LRIS 
system. The spectra were reduced and calibrated using 
standard procedures (e.g., \citealt{matheson00}). Clouds were present 
on the night of 2009 Apr.\ 25, and the observing setup from that night 
did not include any wavelength overlap between the red and blue arms of the 
Kast spectrograph; thus, the absolute flux calibration on this night is 
less certain than on the other nights. For the Kast spectra, we observed 
the SN with the slit placed at the parallactic angle, so the relative 
spectral shapes should be accurate \citep{fil82}. LRIS is equipped with 
an atmospheric dispersion corrector; nevertheless, we observed the SN at 
the parallactic angle on this night as well. In Fig.~\ref{spectra} 
we show the spectral sequence of \sn, and a summary of our observations 
is given in Table~\ref{speclog}. 

\begin{table}
\begin{minipage}{80mm}
  \centering
\caption{Log of spectroscopic observations.}
\begin{tabular}{cccc}
\hline
Epoch\footnote{Defined as rest-frame days relative to day 0, 2007 Sep.\ 13.}
 & UT Date & Telescope/ & Exposure \\
 & & Instrument & (s) \\
\hline
560 & 2009-04-18.458 & Shane 3-m/Kast & 1500 \\
567 & 2009-04-25.479 & Shane 3-m/Kast & 1500 \\
652 & 2009-07-23.274 & Shane 3-m/Kast & 1800 \\
711 & 2009-09-22.236 & Keck I 10-m/LRIS & 600 \\
\hline
\end{tabular}
\label{speclog}
\end{minipage}
\end{table}

\subsection{Host Galaxy}

There is no catalogued redshift for the host galaxy of \sn. To determine 
the redshift of the SN we examined the LRIS spectrum for the 
presence of narrow emission lines. Based on a fit to the 
[O~III] $\lambda\lambda$4959, 5007 doublet we adopt a heliocentric 
redshift of $z = 0.0411$ for \sn\ (not corrected for peculiar motions
or Virgo infall). This corresponds to a luminosity 
distance of $d_L = 179$ Mpc ($H_0 = 71$ \kms Mpc$^{-1}$, $\Omega_M = 0.27$, 
$\Omega_\Lambda = 0.73$), which we adopt throughout the remainder of this 
paper. 

The reddening of \sn\ by the host galaxy is uncertain. In our Kast spectra 
we do not detect any absorption from Na~I D, and 
therefore we adopt $A_{\lambda, {\rm host}} = 0$ mag.

We confirm the possible detection of the host galaxy by \citet{catelan2009} 
in a stack of SDSS $g$, $r$, and $i$-band images of the field. When the 
$g$ + $r$ + $i$ stack is calibrated to the SDSS $r$ band, we find that the 
host has $r$ = 22.75 $\pm$ 0.15 $\pm$ 0.08 mag in a 2.2\arcsec\ diameter 
aperture, where the two errors represent the statistical and calibration 
uncertainties, respectively.

As first noted by \citet{catelan2009}, the field of \sn\ was observed by the 
Galaxy Evolution Explorer (GALEX; \citealt{martin2005}). On 2004 May 17, the 
field was imaged as part of the all-sky imaging survey (AIS); the host was 
not detected to a 5$\sigma$ limiting magnitude of $FUV \ga$ 20.2 and 
$NUV \ga$ 20.5 mag. The field was reimaged as part of the medium imaging 
survey on 2008 Jun.\ 05/06, and \sn\ was detected at $FUV = 21.175 
\pm 0.094$ mag and $NUV = 19.929 \pm 0.036$ mag.

\section{Results}

\subsection{Photometric Analysis}

At $z$ = 0.0411, the absolute peak magnitude of \sn\ was a 
relatively modest $M_r \approx -19.1$ mag. This places \sn\ well below 
the most luminous SNe~IIn, such as SN~2006gy ($M_R = -21.7$ mag; 
\citealt{ofek06gy, smith07-2006gy}) and SN~2008fz ($M_V = -22.3$ mag; 
\citealt{drake09-2008fz}), and more in the range of the well-studied 
SN~IIn~1988Z ($M_R \la -18.9$ mag; \citealt{turatto1993, stathakis1991}). 
Assuming no bolometric correction, the 
total integrated optical output from \sn\ during the first $\sim$700 days 
after discovery is $\sim$2 $\times 10^{50}$ erg. Between day 550 and 720 
the SN declines in the optical and the $J$ band, while it actually gets 
brighter in the $H$ and $K_s$ bands. During this time, the linear decay 
rates are as follows: $\beta_B$ = 0.51 mag (100 day)$^{-1}$, 
$\beta_V$ = 0.50 mag (100 day)$^{-1}$, $\beta_R$ = 0.34 mag (100 day)$^{-1}$, 
$\beta_I$ = 0.43 mag (100 day)$^{-1}$, $\beta_J$ = 0.30 mag (100 day)$^{-1}$, 
$\beta_H$ = $-$0.03 mag (100 day)$^{-1}$, and $\beta_{K_s}$ = $-$0.19 mag 
(100 day)$^{-1}$.
These decline rates, all slower than the expected rate of decline for 
radioactive $^{56}$Co, 0.98 mag (100 day)$^{-1}$, strongly suggest that 
the SN is still being powered by CSM interaction $\sim$700 days after 
explosion.

For comparison purposes, in addition to the light curve of \sn, 
Fig.~\ref{long-lc} shows the light curve of the standard Type II-P 
SN~1999em \citep{leonard02-99em} as it would have appeared at the redshift 
of \sn. As well as illustrating the very long rise time of \sn, this 
comparison shows that \sn\ was fairly luminous at the time of discovery. 
Thus, despite the sparse sampling over the first $\sim$200 days, this 
large luminosity suggests that the early detections are not related to 
a pre-SN outburst, as was observed for SN~2006jc \citep{foley06jc,
pastorello07-06jc}. CRTS images indicate that the SN was rising over 
the first $\sim$400 days \citep{mahabal2009}, which provides further 
evidence against a pre-SN outburst. Fig.~\ref{long-lc} also shows the 
light curve of the long lived, 
interacting, Type IIn SN~1988Z \citep{turatto1993}. The explosion date of 
SN~1988Z is not well constrained (see Section~4.3), so we shift the first 
epoch of detection to day~0. Notice the similarity of the decline rate 
of \sn\ and SN~1988Z around day $\sim$600; these SNe also have very similar 
late-time spectra (see Section~4.2).

We were unable to fit the UV through NIR spectral energy distribution (SED) 
of \sn\ to a single-temperature blackbody model. This is not surprising, 
as SNe~II are dominated by emission lines at late times, and 
single-temperature blackbody models typically apply only to young SNe~II 
(see, e.g., \citealt{filippenko1997}). Direct integration of the UV$-$NIR 
SED shows that the bolometric correction, relative to the $I$ band, is a 
factor of $\sim$3 in luminosity, corresponding to $\sim$1.2 mag, on day 
$\sim$560. Some features stand 
out from the SED: there is a strong $R$-band excess relative to the 
other optical bands, with $V-R = 0.8$ mag and $R-I = -0.2$  mag on day
715. The red $V-R$ colour relative to a blue $R-I$ colour can be 
attributed to the H$\alpha$ emission with large equivalent 
width. This emission also accounts for the relatively 
slow decay rate in the $R$ band, as compared to the other optical bands. 
There is also a NIR excess relative to 
the $I$-band flux. In fact, this excess increases with time, $I-K \approx 2.8$ 
mag on day $\sim$560 and $I-K \approx 3.7$ mag on day $\sim$710, which is 
indicative of the growing importance of dust in the emission from 
\sn. 

A NIR excess at late times has been observed in many 
SNe~IIn (e.g., \citealt{gerardy2002}), and it is the result of either 
new dust formation (e.g., SN~2005ip; \citealt{smith09-2005ip, fox2009})
or the presence of a NIR light echo from preexisting dust \citep{dwek1983}, 
or both. To distinguish between these two possibilities, which are 
virtually identical from photometry alone, requires well-sampled 
optical spectra (see Section 3.2.2) since line profiles are expected to 
change with time if new dust is being formed. On day 706, corresponding to 
our last PAIRITEL observation, the $H-K_s$ colour of \sn\ was $\sim$1.2 mag. 
Assuming that the dust radiates as a perfect blackbody, this colour 
corresponds to a dust temperature of $T_{\rm dust}~\approx~1320$ 
K, while the $K_s$-band measurement corresponds to a dust luminosity 
$L_{\rm NIR}~\approx~4.8 \times 10^{42}$ erg s$^{-1}$, assuming no 
bolometric correction as we cannot constrain the emission in the mid-IR. 
This luminosity is very large, though upon making similar assumptions 
\citet{gerardy2002} found NIR luminosities $> 10^{42}$ erg s$^{-1}$ for 
SNe~1995N and 1997ab at late times. Gerardy~et~al. also found that the NIR 
luminosity was roughly an order of magnitude greater than the X-ray emission 
from SN~1995N, which is also the case for \sn\ (see Section~3.3). The 
$H-K_s$ colour of \sn\ is bluer 
than those of the Gerardy~et~al.\ sample at a similar epoch, which may 
be a result of the long rise of \sn\ or a contribution to the NIR 
emission from the SN in addition to the dust. Future NIR observations, as 
the underlying SN light continues to fade, will place more stringent 
constraints on the dust near \sn.

\subsection{Spectroscopic Analysis}

Our spectra of \sn\ at ages $>$560 days are very similar to those 
of an unpublished spectrum, from our spectral database, of the Type IIn 
SN 1988Z taken at a comparable epoch, as shown in 
Fig.~\ref{spectra}. The general features resemble those of 
several other SNe~IIn \citep{filippenko1997}: there are 
prominent Balmer and He~I emission lines, most notably 
the dominant H$\alpha$ emission, which primarily feature intermediate-width 
emission components with full width at half-maximum intensity (FWHM) 
$\sim$1650 \kms\ in the case of \sn. The higher-resolution Keck spectrum 
reveals a number of narrow, marginally resolved (FWHM $\la$ 170 km s$^{-1}$) 
emission  lines, including H$\alpha$, H$\beta$, [O~III] $\lambda\lambda$4363, 
4959, 5007, and He~I $\lambda$7065. The only high-ionisation line we 
detect is [Fe~X] $\lambda$6375.
The relative lack of narrow forbidden lines and the low intensity ratio of 
[O~III] $\lambda\lambda$4959, 5007 to [O~III] $\lambda$4363 
suggest a large electron density for the 
ejecta \citep{filippenko+halpern1984,filippenko89-1987F,stathakis1991}. 
The relative spectral features do not show strong evolution between 
days 560 and 711.

We can estimate the density of the unshocked emitting material based on the 
relative intensities of the three [O~III] lines mentioned above. 
Note that [O~III] 
$\lambda$4363 is only resolved in our day 711 Keck spectrum, so 
the following estimate of the density is at that epoch only. 
Following a removal of the 
underlying continuum, we fit three Gaussian profiles to [O~III] 
$\lambda$4363 and [O~III] $\lambda\lambda$4959, 5007, and find that 
${\cal R} = I[\lambda4959 + \lambda5007]/I[\lambda4363] \approx 1.7$. 
As mentioned above, the strong presence of [O~III] $\lambda$4363 indicates a high 
density for the emitting material. In fact, with ${\cal R}$ $\approx$ 1.7, 
the electron density, $n_{\rm e}$, 
must be $> 10^6$ cm$^{-3}$ regardless of the temperature of the emitting 
gas. Typically, [O~III] emission comes from photoionised regions with 
$T = 16,000$--20,000, in which case $n_{\rm e} \approx 10^7$ cm$^{-3}$ (see 
Fig.~11 in \citealt{filippenko+halpern1984}). Note that this 
high-density material is likely only present within clumps in the CSM (see 
Section~4.2), and that interclump portions of the CSM have a 
lower density.

\subsubsection{The H$\alpha$ Profile}

\begin{figure}
\includegraphics[width=84mm]{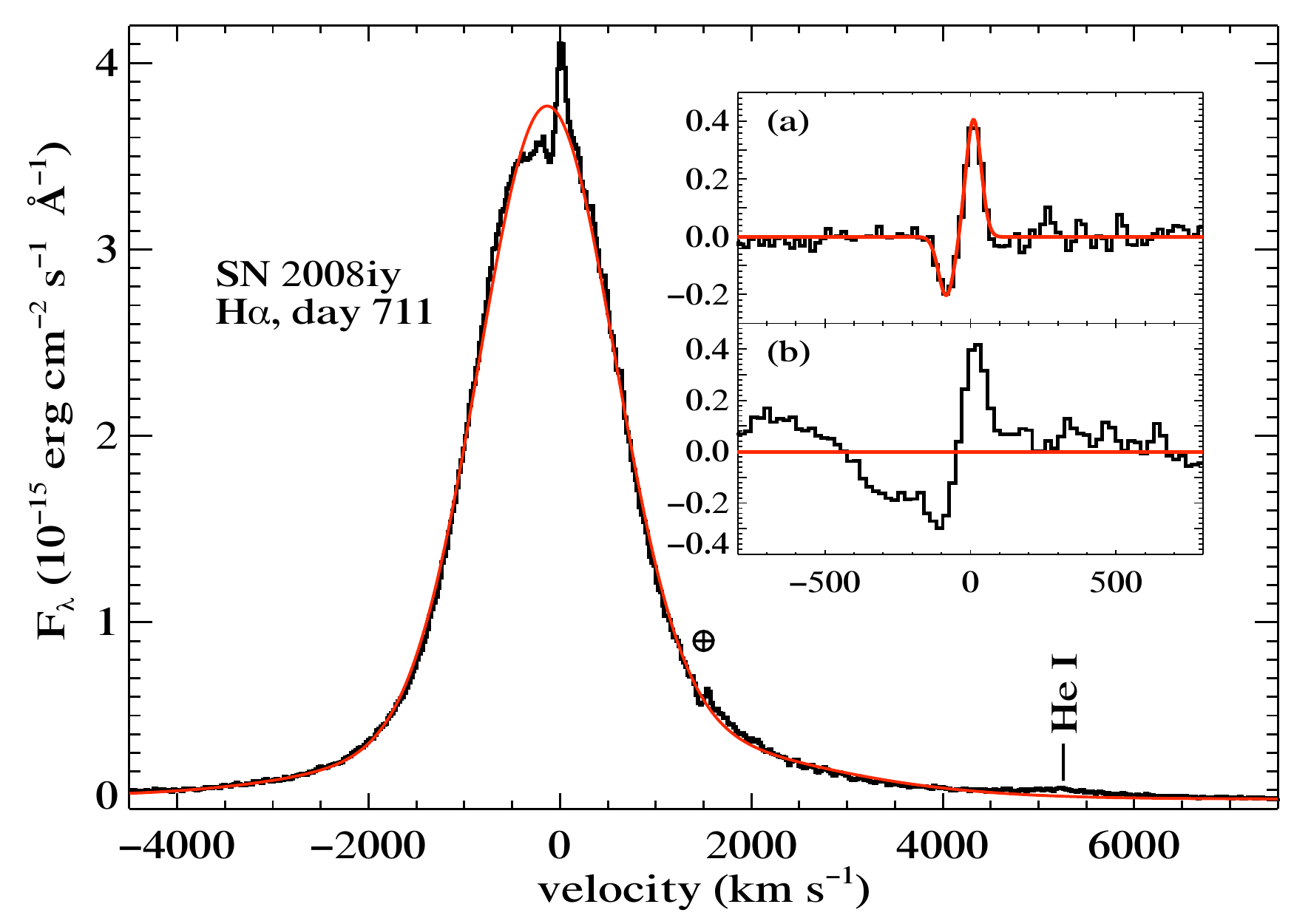}
\caption{Detailed view of the H$\alpha$ profile from the Keck spectrum taken on 
day 711. {\it Main panel}: The black line shows the observed H$\alpha$ 
profile, while the smooth red line illustrates a two-Gaussian fit (intermediate, 
FWHM $\approx 1650$ \kms; broad, FWHM $\approx 4500$ \kms)
 to the profile. Residual emission and absorption from a narrow 
component to the profile can easily be seen. The location of 
He~I $\lambda$6678 has been marked, and the residuals from an imperfect 
telluric absorption correction have been marked by the $\oplus$ symbol. 
{\it Inset (a)}: Close-up view of the  
residual, narrow P Cygni component. The smooth red line shows a
fit to the feature that includes a Gaussian in emission and a Gaussian 
in absorption (both Gaussians have FWHM $\approx$ 75 \kms). The absorption 
minimum of the narrow absorption is located at $-100$ \kms. Note that for this 
fit we removed the broad plus intermediate components using a spline fit 
over the region from $-1000$ \kms\ to $1000$ \kms (see text). {\it Inset (b)}: 
Residuals following the subtraction of the two-Gaussian broad plus intermediate 
fit from the observed H$\alpha$ profile. The absorption minimum still 
occurs at $-100$ \kms, but the BVZI extends to $\sim -450$ \kms, 
meaning the progenitor wind speed may be $>100$ \kms\ (see text). 
}
\label{Halpha}
\end{figure}

As seen in Fig.~\ref{spectra}, the H$\alpha$ emission feature 
dominates over all the other lines. In the high-resolution Keck 
spectrum, we see evidence for three distinct emission features: 
a broad component (FWHM $\approx$ 4500 km s$^{-1}$), an intermediate 
component (FWHM $\approx$ 1650 km s$^{-1}$), and a marginally 
resolved, narrow P Cygni feature (FWHM $\approx$ 75 km 
s$^{-1}$), as shown in Fig.~\ref{Halpha}. The main panel of 
Fig.~\ref{Halpha} shows a fit to the H$\alpha$ profile (red line) 
which includes a broad (FWHM $\approx$ 4500 km s$^{-1}$) and an intermediate 
(FWHM $\approx$ 1650 km s$^{-1}$) component. Panel (a) in Fig.~\ref{Halpha} 
shows a close-up view of 
the narrow emission after the broad and intermediate components have been 
removed using a spline fit. We fit the spline to the observed H$\alpha$ 
profile from $-1000$ \kms\ to $1000$ \kms\ after masking 
the region between $-200$ \kms\ and $200$ \kms. A narrow absorption 
minimum such as this, located at $\sim -$100 km s$^{-1}$, has been seen 
in a number of typical SNe~IIn (e.g., SNe 1997ab, 1997eg, and 
1998S, \citealt{salamanca02-97eg,salamanca2000-97ab,fassia2001}), and 
it traces the outflow velocity of the progenitor wind. We note that while 
the narrow absorption is only marginally resolved, a P Cygni profile 
with characteristic speeds of $\sim$10 \kms, typical of red supergiants 
(RSGs), would go completely unresolved in the Keck spectrum. Therefore, 
this P Cygni profile must be associated with an outflow velocity that 
is $>$10 \kms, which will have important consequences for the progenitor 
(see Section~4.3).

It is interesting that the broad plus intermediate fit overestimates the 
H$\alpha$ flux out to about $-$450 km s$^{-1}$. In panel (b) of 
Fig.~\ref{Halpha} we show the residual flux, centred on H$\alpha$, 
following the subtraction of our two-Gaussian fit to the 
broad plus intermediate emission. The blue velocity at 
zero intensity (BVZI) extends to roughly $-450$ \kms, which means that
the true wind speed from the progenitor, as traced by the absorbing gas 
moving directly along the line of sight toward the SN, may be $>$100 \kms\ 
and possibly as high as $\sim$450 \kms. If this feature represents a 
true lack of emission, it would be evidence for a two-component CSM: 
one with velocity $v_{\rm w,1} \approx 100$ km s$^{-1}$ and the other with 
BVZI $\approx$ $v_{\rm w,2}$ $\approx 450$ km s$^{-1}$. 
Similar features were seen in SN 1998S \citep{fassia2001}, which were
modelled by \citet{chugai2002} to be a slow wind accelerated to higher 
velocities by radiation from the SN photosphere, and in SN 2006gy, 
which \citet{smith09-2006gy} argued was the result of a CSM shell that 
had been ejected from the progenitor in a pre-SN eruption. These two 
scenarios predict different behaviour as the SN evolves. 

In the radiatively accelerated scenario, as shown by \citet{chugai2002} 
for SN~1998S, the second, faster component to the wind has a negative 
velocity gradient. 
Observationally, this effect manifests itself as a BVZI 
for the second, faster CSM component that decreases with time, as observed 
in SN~1998S \citep{fassia2001}. In the shell-ejection scenario, the wind 
exhibits a positive velocity gradient as the ejected shell is freely 
expanding, which is thought to be the result 
of an explosive ($\sim 10^{49}-10^{50}$ erg) mass-loss event (see e.g., 
\citealt{chugai94w}). This scenario leads to a BVZI that increases 
with time, as 
was observed for SN~2006gy \citep{smith09-2006gy}. With only a single 
high-resolution spectrum, we are unable to probe the evolution of this 
feature, but we note that the velocity of a radiatively accelerated CSM 
is proportional to $t^{-2}$ \citep{fransson96-1993J}, so this mechanism is 
unlikely to be significant at late times ($>$ a few hundred days).

\begin{figure}
\includegraphics[width=84mm]{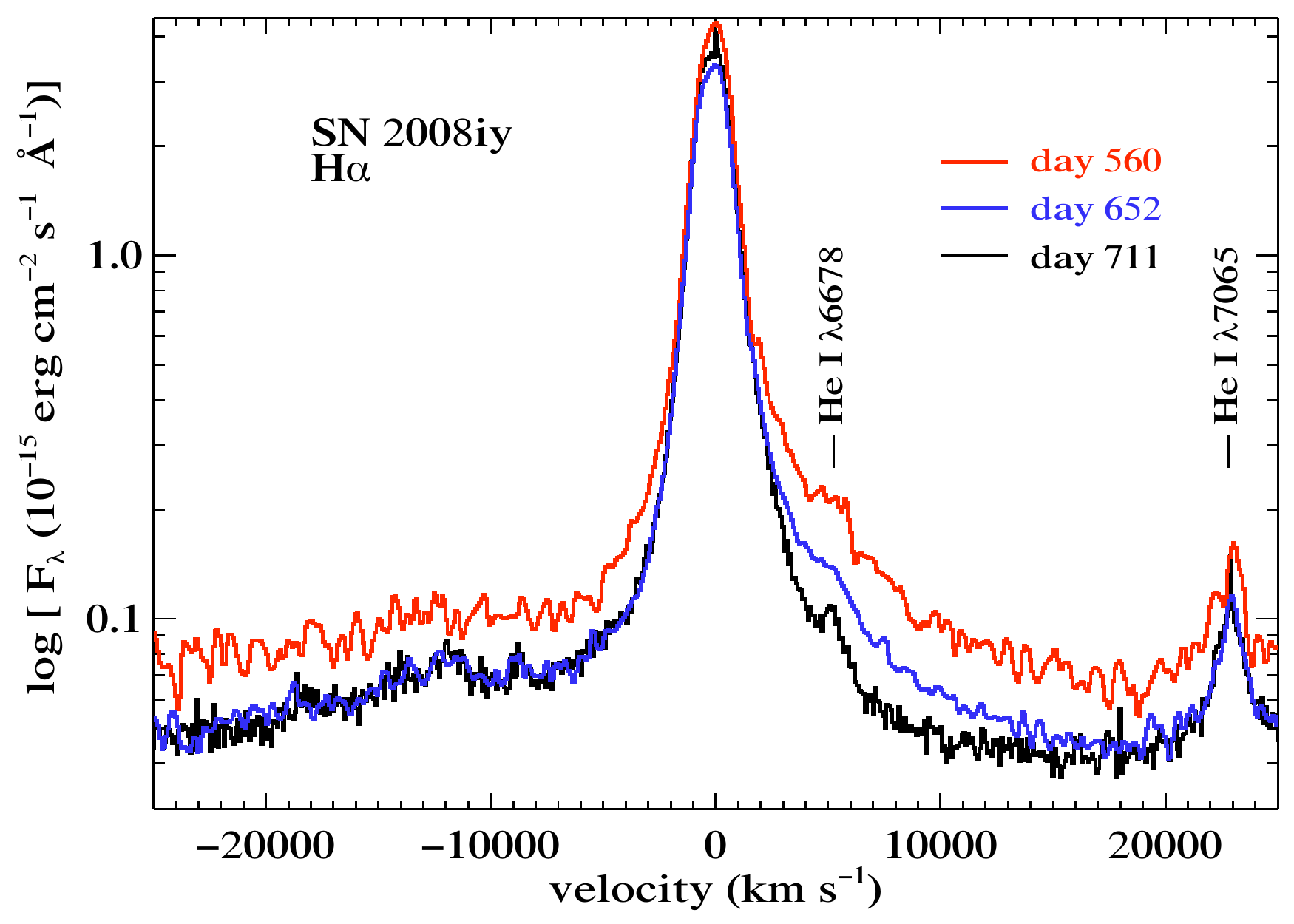}
\caption{Evolution of the H$\alpha$ profile of \sn\ between day 560 and 
day 711. The profile remains largely unchanged, with the exception of a 
variable spectral feature between $\sim$200 \kms\ and 
20,000 \kms, from day 652 to day 711. Similar spectral features 
have been observed in other interacting SNe (see text). Emission lines of
He~I $\lambda$6678 and He~I $\lambda$7065 are labelled. 
}
\label{Halpha-evolve}
\end{figure}

As does the overall continuum, the H$\alpha$ luminosity declines over the 
course of our observations. In Table~\ref{tab-halpha} we summarise the 
observed properties of the H$\alpha$ profile from our spectra on days 
560, 652, and 711. We do not include the spectrum from day 567, as it 
was taken under cloudy conditions and the overall flux scaling is 
uncertain, but we do not expect significant spectral evolution between days
560 and 567. The decline in the H$\alpha$ luminosity is accompanied by 
a rise in the equivalent width over this same time period.  
For the narrow emission on day 711, we measure $L({\rm H}\alpha) = (2.8 
\pm 0.6) \times 10^{39}$ erg s$^{-1}$, though we note that this value is 
likely underestimated because the blue emission wing is probably being 
partially absorbed. 
The FWHM of the intermediate component is $\sim$1650 km s$^{-1}$ at all 
epochs, while the FWHM of the broad component drops from $\sim$5800 \kms\ 
on day 560, to $\sim$5200 \kms\ on day 650, to $\sim$4400 \kms\ on 
day 711. We attribute this change in the broad component to rapid 
evolution of the spectrum redward of H$\alpha$ between days 560 and 711, 
as shown in Fig.~\ref{Halpha-evolve}. 
The nature of this feature is not currently understood: it could 
potentially be an absorption feature, or it could be related to a lack of 
emission from some unidentified species. Similar spectral evolution 
redward of H$\alpha$ has been seen in the interacting SNe 2005gj and 
2006gy \citep{prieto05gj,smith09-2006gy}. We do not associate this feature 
with H$\alpha$, as the red velocity at zero intensity corresponds to 
velocities of $\ga$ 22,000 \kms, and there are no other emission lines that 
show evidence for velocities this large. 

\begin{table*}
\begin{minipage}{140mm}
  \centering
\caption{H$\alpha$ line widths, luminosity, equivalent width, and ratio to H$\beta$.} 
\begin{tabular}{ccccccc}
\hline
Epoch\footnote{Defined as rest-frame days relative to day 0, 2007 Sep.\ 13.}
 & Br. FWHM & Int. FWHM & Nar. FWHM & $L$(H$\alpha$)\footnote{Uncertainty $\sim \pm 10$\%.} 
& EW\footnote{Uncertainty $\sim \pm 10$\%.} & H$\alpha$/H$\beta$ \\
 & (\kms) & (\kms) & (\kms) & (erg s$^{-1}$) & (\AA) & \\
\hline
560 &  5774 & 1641 & ... & $7.86\times10^{41}$ & 2122 & 9.49 \\
652 &  5162 & 1685 & ... & $6.10\times10^{41}$ & 2386 & 12.22 \\
711 &  4405 & 1639 & 78 & $6.50\times10^{41}$ & 2786 & 12.54 \\
\hline
\end{tabular}
\label{tab-halpha}
\end{minipage}
\end{table*}

\subsubsection{He~I Emission}

In addition to the prominent Balmer emission lines, \sn\ exhibits several 
He~I lines in emission, including $\lambda\lambda$3820, 4026, 4471, 5016, 
5875, 6678, and 7065. None of these lines show evidence for a broad component, 
and the intermediate component, FWHM $\approx$ 1100--1400 \kms, is slightly 
narrower than that observed for H$\alpha$. As previously mentioned, He~I 
$\lambda$7065 has a narrow, marginally resolved emission component in 
addition to the intermediate component. 

A systematic, increasing blueshift 
in the He~I profiles has been argued as evidence for dust formation in 
the cool, dense shell formed in the post-shock gas of the Type~IIn SNe 
1998S and 2005ip  \citep{pozzo2004,smith09-2005ip}. Newly formed 
dust in the post-shock region absorbs light from the  
receding ejecta, resulting in a suppression of flux from the red side 
of the emission line. Using He~I $\lambda$7065, the strongest He line in 
our spectra, we searched for changes in the red wing of the profile. The 
Kast spectra from days 560 and 567 are very noisy, but in the higher 
signal-to-noise ratio spectra from days 652 and 711 we 
see no evidence for a change 
in the red wing of the emission profile. This hints that the observed NIR 
excess (see Section~3.1) is the result of a NIR echo, rather than newly 
formed dust, though we caution that with only two spectra separated by 
$\sim$50 days the lack of change in the He~I profile does not constitute 
definitive evidence for this.

\subsection{X-ray Analysis}

To determine the X-ray luminosity of \sn\ we assume a thermal plasma 
spectrum with temperature 0.6 keV (see, e.g., \citealt{immler2005}), which has 
been absorbed by a Galactic column density of $n({\rm H}) = 4.8 \times 
10^{20}$ cm$^{-2}$ \citep{kalberla2005}. Using the 0.3--10 keV XRT count 
rate (see Section~2.1) and 
PIMMS\footnote{{http://heasarc.nasa.gov/Tools/w3pimms.html.}}, we measure 
the unabsorbed flux, assuming no absorption within the host, to be 
(6.1 $\pm$ 1.9) $\times$10$^{-14}$ erg cm$^{-2}$ s$^{-1}$. At the adopted 
distance of \sn\ this corresponds to a luminosity of $L_{\rm X} = 
(2.4 \pm 0.8) \times 10^{41}$ erg s$^{-1}$. Note that \citet{fox2000} fit the 
late-time ($>$ 3 yr) X-ray spectra of SN~1995N with a larger temperature 
of 9.1 keV, which would correspond to a flux and luminosity of 
(9.1 $\pm$ 2.9) $\times$10$^{-14}$ erg cm$^{-2}$ s$^{-1}$ and 
$(3.5 \pm 1.1) \times 10^{41}$ erg s$^{-1}$, respectively, were we to adopt 
this value for \sn. Similar luminosities ($L_X > 10^{41}$ erg s$^{-1}$) are 
observed for other interacting SNe more than 1 year post explosion 
(e.g., SNe 1988Z and 1995N; \citealt{fox2000}).

\section{Discussion}

\subsection{Mass Loss in the \sn\ Progenitor}

We assume that the continuum emission traced by the light curve of \sn\ 
is powered solely by CSM interaction. The half life of $^{56}$Ni is 
$\tau \approx$  6.0 days, much too short to power the $>$ 400 day rise of \sn, 
while the photometric decay rates are too slow to be powered by $^{56}$Co 
(see Sec. 3.1), implying that radioactivity contributes minimally to the light 
curve of \sn. Furthermore, the SN ejecta will suffer considerable adiabatic 
losses during the 400 day rise of the light curve, suggesting that 
interaction is the only 
viable mechanism for powering \sn\ over this entire time. 

Following our assumption of interaction-driven luminosity, the 
expelled gas from the progenitor overtaken by the shock at each epoch can be 
written as (see \citealt{chugai1994})

\begin{equation}
\dot{M} = \frac{2}{\psi} L \frac{v_{\rm w}}{V_{\rm SN}^{3}},
\label{optical-mdot}
\end{equation}
where \mdot\ is the mass-loss rate of the progenitor, $\psi$ is an 
efficiency factor describing the conversion of kinetic energy into 
radiation, $L$ is the luminosity of the SN, $v_{\rm w}$ is the wind speed 
of the progenitor, and $V_{\rm SN}$ is the velocity of the blast wave 
overrunning the CSM. We adopt $\psi$ = 0.5, but this number 
is likely an overestimate given the optically thin nature of the emission 
(see Section 4.2). The CSM wind is only probed by our higher-resolution 
spectrum from day 711, so we adopt $v_{\rm w}$ = 100 \kms\ based on the 
narrow P Cygni absorption seen in H$\alpha$, though this may not be valid 
at all times between days 0 and 700 (see below). We adopt $V_{\rm SN}$ = 
5000 \kms\ based on the typical widths of the broad H$\alpha$ 
emission, which should trace the speed of the blast wave. Unlike our 
assumption of a constant wind speed, the adoption of a constant-velocity 
blast wave at all times should at the very least be valid after the peak, 
at which point the CSM density is decreasing with increasing radius. 
Hydrodynamic modelling of SNe~IIn demonstrates that the blast wave is 
quickly decelerated to a roughly constant expansion speed as the ejecta 
sweep up successively more CSM (see, e.g., \citealt{chugai94w}). This 
scenario may not apply to \sn, however, because the $\sim$400 day rise 
time implies that the CSM density may have been increasing with radius, in 
which case the blast wave may have been continually decelerated 
during the rise. 

Following the above assumptions we can use Equation~\ref{optical-mdot} 
to determine the mass loss traced by the optical luminosity at a number 
of different epochs of interest. At each epoch we determine the radius, 
$R = V_{\rm SN}t_{\rm SN}~\approx~5000$ \kms\ $t_{\rm SN}$, and the 
luminosity based on the light curve shown in Fig.~\ref{long-lc}. We do 
not adopt the bolometric correction determined in Section~3.1 for two 
reasons: (i) it is unclear whether this same correction is valid at 
early times, and (ii) emission from a NIR echo should not be incorporated 
into Equation~\ref{optical-mdot}, as that contribution to the total 
luminosity is not directly the result of CSM interaction. Thus, the 
estimates below of the luminosity do not reflect any emission from dust. 
At early times, day $\sim$153, $L \approx 6.8 \times 10^{8}$ \lsun, which 
corresponds to \mdot\ $\approx$ 1.3 $\times~10^{-2}$ \msun\ yr$^{-1}$ at 
$R \approx 6.6 \times 10^{15}$ cm, 
while at the time of peak $L \approx 1.2 \times 10^{9}$ \lsun, 
$R \approx 1.7 \times 10^{16}$ cm, and \mdot\ $\approx$ 2.3 
$\times~10^{-2}$ \msun\ yr$^{-1}$. On day 563, coincident with 
our X-ray observations of \sn\, the optical luminosity was $L \approx 7 
\times 10^{8}$ \lsun, which corresponds to $R \approx 2.4 \times 10^{16}$ cm 
and \mdot\ $\approx$ 1.3 $\times~10^{-2}$ \msun\ yr$^{-1}$. On day 714, 
roughly coincident with our high-resolution Keck spectrum, 
$L \approx 4 \times 10^{8}$ \lsun, $R \approx 3.0 \times 10^{16}$ cm, 
and \mdot\ $\approx$ 0.8 $\times~10^{-2}$ \msun\ yr$^{-1}$.

These results are quite remarkable; they suggest that at a time $t \approx 
V_{\rm SN}t_{\rm SN}/v_{\rm w} \approx $ 5000 \kms\ $\times$ 400 day/100 \kms\ 
$\approx$ 55 years prior to the explosion of \sn, the progenitor underwent a 
period of heightened mass loss. This enhanced mass-loss period was then 
followed by a period of decreasing mass loss leading up to the time of the SN 
explosion. While the quantitative results for the mass loss discussed above 
are sensitive to our adopted quantities, most specifically the 
progenitor wind speed and the speed of the SN blast wave,
the fact remains that the continuum luminosity did increase over a period 
of $\sim$400 days after the SN explosion. Rearrangement of 
Equation~\ref{optical-mdot} shows that this luminosity increase is 
proportional to the wind-density parameter, $w = \dot{M}/v_{\rm w}$. 
Thus, $w$ must have {\it increased} over a distance 
$R \approx V_{\rm SN}t_{\rm SN} \approx 1.7\times 10^{16}$ cm from the 
progenitor, regardless of the true wind speed over that 
distance. An increasing value of $w$ means that during roughly the
century prior to the SN, there was a significant change 
in the wind properties of the progenitor. 

Using our X-ray detection of \sn\ we have an alternative method to 
probe the mass-loss history of the progenitor on day $\sim$563. Following 
\citet{immler2005}, the X-ray luminosity may be written as

\begin{equation}
L_{\rm X} = \frac{4}{({\pi}m)^2} \Lambda(T) 
\left(\frac{\dot{M}}{v_{\rm w}}\right)^2 (V_{\rm SN}t)^{-1},
\label{x-ray}
\end{equation}
where $L_{\rm X}$ is the X-ray luminosity, $m$ is the mean mass per particle 
($m$ = 2.1 $\times 10^{-24}$ g for a H+He plasma), $\Lambda(T)$ is the 
cooling function of a plasma heated to temperature $T$, \mdot\ is the 
mass-loss rate of the progenitor, $v_{\rm w}$ is the wind speed of the 
progenitor ($\sim 100$ \kms, see above), $V_{\rm SN}$ is the speed of the 
SN blast wave ($\sim 5000$ \kms, see above), and $t$ is the time since 
explosion, $\sim$563 days. Assuming a temperature $T = 10^7$ K, appropriate 
for an optically thin thermal plasma \citep{immler2005}, the effective 
cooling function is $\Lambda(T) = 3 \times 10^{-23}$ erg cm$^{3}$ s$^{-1}$. 
The X-ray luminosity on day 563 was $L_{\rm X} \approx 2.4 \times 10^{41}$ erg 
s$^{-1}$. Substituting these values into Equation~\ref{x-ray}, we find 
\mdot\ $\approx 0.7 \times 10^{-2}$ \msun\ yr$^{-1}$, which shows agreement 
with the value derived from the optical continuum, $\sim 1.3 \times 10^{-2}$ 
\msun\ yr$^{-1}$, given the uncertainty in a number of the assumptions 
we have adopted.

An additional estimate of the mass-loss rate of the progenitor comes 
from the narrow H$\alpha$ emission seen in our Keck spectrum on 
day 711. \citet{chugai95g} show that the H$\alpha$ emission from the 
unshocked CSM is related to $w$:

\begin{equation}
L({\rm H}\alpha) = \frac{1}{4\pi{r_1}} \alpha_{32} h\nu_{23}(xXwN_{\rm A})^2
\left(1 - \frac{r_1}{r_2}\right), 
\label{halpha-mdot}
\end{equation}
where $r_1$ is the inner radius corresponding to the position of the 
SN blast wave, $r_2$ is the outer radius related to the fast-moving 
forward shock, $\alpha_{32}$ is the effective recombination 
coefficient for H$\alpha$, $h\nu_{23}$ is the energy of an H$\alpha$ photon, 
$x$ is the degree of H ionisation, $X$ is 
the hydrogen mass fraction, and $N_{\rm A}$ is Avogadro's number. We cannot 
constrain $x, X,$ or $r_{2}$, so we conservatively adopt $x = 1$, $X = 1$, 
and $r_2 \gg r_1$, all of which have the effect of minimising $w$. Assuming 
Case B H$\alpha$ recombination, $\alpha_{32} = 8.64 \times 10^{-14}$ cm$^3$ 
s$^{-1}$, which is appropriate assuming that the narrow emission comes 
from photoionised gas. 
Substituting the narrow H$\alpha$ luminosity ($L({\rm H}\alpha) 
= 2.8 \times 10^{39}$ erg s$^{-1}$) and $r_1 = V_{\rm SN}t_{\rm SN} = 
3.0 \times 10^{16}$ cm into Equation~\ref{halpha-mdot}, 
we find that $w \approx 10^{17}$ g cm$^{-1}$. With $v_w = 100$ 
\kms\ on day 711, this corresponds to \mdot\ $\approx 1.6 \times 10^{-2}$ 
\msun\ yr$^{-1}$. At 
a similar epoch (see above), we find \mdot\ $\approx 0.8 \times 10^{-2}$ 
\msun\ yr$^{-1}$ based on the optical luminosity. Again, the agreement 
between these two methods to within a factor of $\sim$2 is 
reasonable given the uncertainties in our adopted parameters. 

\subsection{Late-Time Emission and the Similarity to SN 1988Z}

To explain the late-time (i.e., post peak) emission from \sn\ we adopt 
a model that is virtually identical to that 
developed by \citet{chugai1994} for SN~1988Z: the 
CSM contains optically thick clumps in addition to a rarefied wind 
between these clumps (see, e.g., Fig. 1 of \citealt{chugai1994}). As the 
SN ejecta sweep through the CSM, they drive a 
fast-moving forward shock into the rarefied wind. This shocked material 
forms a cool dense shell between the forward and reverse shocks, and 
gives rise to the broad emission component seen in H$\alpha$ (FWHM $\approx$ 
5000 \kms\ for \sn). At the same time a slower shock 
is being driven into the dense clumps, which leads to the intermediate-width 
emission (FWHM $\approx$ 1650 \kms\ in the case of \sn) seen in the 
spectra. The narrow P Cygni profile results from the pre-shock 
photoionised wind. 

The observed similarities between \sn\ and SN~1988Z 
provide further evidence that comparable models are appropriate for the 
two SNe. In addition to having similar spectra at late times (see 
Fig.~\ref{spectra}), SNe~1988Z and 2008iy have large X-ray 
luminosities (both SNe have $L_{\rm X} \ga 10^{41}\ {\rm erg s}^{-1}$; 
\citealt{fabian88z}, this work), and similar late-time decline rates 
($\sim$0.3$-$0.5 mag (100 day)$^{-1}$ in the optical; \citealt{turatto1993}, 
this work), which in both cases are slower than the expected bolometric 
decline of radioactive $^{56}$Co. Estimates for the mass-loss rate from 
the progenitor of SN~1988Z vary by roughly an order of magnitude. From 
X-ray observations \citet{schlegel2006} find $\dot{M}_{\rm 1988Z} \approx 
10^{-3}$ \msun\ yr$^{-1}$, while \citet{williams2002} estimate 
$\dot{M}_{\rm 1988Z} \approx 10^{-4}$ \msun\ yr$^{-1}$ based on radio 
observations, and the models of the optical emission by \citet{chugai1994} 
yield $\dot{M}_{\rm 1988Z} \approx 7 \times 10^{-4}$ \msun\ yr$^{-1}$. While 
these estimates are 1--2 orders of magnitude less than those for 
\sn, we note that in each of the above estimates for $\dot{M}_{\rm 1988Z}$ 
a wind speed of 10 \kms\ was adopted for the progenitor 
of SN~1988Z. If SN~1988Z had a progenitor wind speed closer to $\sim$100 \kms, 
a very reasonable possibility given the wind speed of \sn\ and other 
similarities between the two SNe, this would result in an increase in the 
estimated mass-loss for the progenitor by a factor of $\sim$10, thereby 
bringing the estimates for the two SNe into accord. 

Based on the late-time emission, \sn\ seems to belong to the group of 
SNe~IIn that exhibit a slow evolution sustained by long-lived 
($\ga$ 1 decade) CSM interaction. In addition to SN~1988Z, other members 
of this group include SN~1986J \citep{rupen86J}, SN~1995N 
\citep{fox2000,fransson2002}, and possibly also the VLSN~2003ma 
\citep{rest03ma}. Both SN~1986J and SN~1995N exhibit 
evidence for a clumpy progenitor wind, like \sn\ and SN~1988Z. 
\citet{chugai86J} modelled the 
X-ray emission from SN~1986J as the interaction between the SN ejecta and 
a clumpy wind (though note that \citealt{houck98-1986J} prefer a model 
with a smooth CSM, but they conclude that they cannot rule out the 
clumpy model). The evidence for a clumpy wind from the progenitor 
of SN~1995N comes from both the optical spectra \citep{fransson2002} 
and the X-ray emission \citep{zampieri95N}. Also, the optical decline of 
both SNe~1986J and 1995N is very slow: SN~1986J declined by $\la$ 1 mag 
in the optical between 1994 and 2003 \citep{milisavljevic2008}, while 
the $V$-band decline of SN~1995N was only 2.2 mag between 1998 and 2003 
\citep{zampieri95N}. 

The duration of the interaction means that the large mass-loss rates 
from the respective progenitors must have been sustained for 
at least $\sim 100$ yr prior to core collapse after the conservative 
assumption that $V_{\rm SN}$ = 1000 \kms, and $v_{\rm w}$ = 100 \kms. 
Typically the blast wave continues to expand at $>$ 1000 \kms\ (SN~1988Z 
had a broad component that remained nearly constant at $\sim$2000 \kms\ 
from day $\sim$1500 to 3000; \citealt{aretxaga88z}), which would make 
this time period $>$ 100 yr. These very long-lived SNe~IIn are 
therefore connected in that their progenitors experienced lengthy periods 
with high mass-loss rates. This stands in stark contrast to 
the Type IIn SN~1994W, which had a light curve with an abrupt drop 
$\sim$100 days post explosion \citep{sollerman94w}. The mass loss from 
the progenitor of SN~1994W has been modelled to occur in a short 
($\la$1 yr), violent episode \citep{chugai94w}, and the sudden drop in 
the light curve occurs because the SN ejecta have overtaken the dense 
CSM, thereby halting any interaction luminosity.

\subsection{Origin of the 400 Day Rise Time and Implications for the 
Progenitor}

With a model to account for the late-time emission from \sn, we are still left 
with the puzzle of explaining the 400 day rise time.\footnote{We have no 
spectra of the SN while it was still on the rise, thus we have no way 
of knowing if the SN underwent significant spectral evolution prior to 
day 560.}  This rise in the optical is significantly longer than that 
seen in any other known SN. As previously discussed, this 
scenario is possible if the progenitor underwent a phase of enhanced 
mass loss $\sim$55 years prior to the SN explosion. The optical 
luminosity traces the wind-density parameter, $w$, meaning that during 
the decades prior to explosion either (i) the mass-loss rate declined, 
or (ii) the wind speed increased, or (iii) both. 

Episodic periods of enhanced mass loss, sometimes in the form of a shell 
ejection, have been observed for numerous LBVs (see \citealt{humphreys1994}), 
though the underlying physics of these eruptions is not currently well 
understood (see \citealt{smith+owocki2006}). If the progenitor of \sn\ 
had a giant eruption (similar to that of LBVs) 
$\la$ 1 century prior to explosion, this could result in a density profile 
that peaks $\sim$1.7 $\times 10^{16}$ cm from the progenitor. In the 
clumpy-wind scenario described in Section~4.2, this would mean that the ejecta 
are expanding into a wind where the number density of clumps is increasing 
with radius. Thus, as more and more clumps are overtaken by the ejecta,
the continuum luminosity continues to rise, until after $\sim$400 days 
when the ejecta have reached $\sim$1.7 $\times 10^{16}$ cm, the number 
density of clumps begins to decline and so does the optical luminosity. 
Alternatively, the long rise could result from the ejecta expanding into 
a nonspherical wind, such as a bipolar outflow, though this hypothesis 
would need to be examined with detailed hydrodynamical models.

The inferred mass-loss rate for \sn, \mdot\ $\approx 10^{-2}$ \msun\ 
yr$^{-1}$, is similar to that for the first LBV, P Cygni, during 
its great eruption (\mdot$_{\rm P Cyg}$ $\approx 10^{-2}$ 
\msun\ yr$^{-1}$ during the 1600 AD eruption; \citealt{smith+hartigan2006}). 
Furthermore, the terminal wind speed in the P Cygni nebula is 185 
\kms\ \citep{lamers96-PCyg,najarro97-PCyg}, which is similar to the 
observed BVZI for the narrow blue absorption seen in \sn, corresponding 
to $v_{\rm w} \approx 160$--450 \kms. We illustrate these 
comparisons to show that derived properties of the progenitor wind of 
\sn\ are similar to the giant eruption of a Galactic LBV; we are not 
insinuating that the great outburst from P Cygni is a direct analogue to 
the proposed LBV-like eruption from \sn. Such eruptions are not expected 
from RSGs.

Further evidence that the progenitor of \sn\ could not have had a RSG 
progenitor comes from the observed narrow P Cygni H$\alpha$ 
profile. RSGs have typical wind speeds of $\sim$10 \kms, with extreme 
RSG winds reaching 40 \kms\ (see \citealt{smith07-2006gy}). The 
observed 100 \kms\ wind speed from \sn\ is more characteristic of LBVs (e.g., 
P Cygni, see above) or the escape velocity from a blue supergiant. Similar 
$\sim$100 \kms\ P Cygni profiles have been seen in a number of 
SNe~IIn, such as SNe~1997ab \citep{salamanca2000-97ab}, 
1997eg \citep{salamanca02-97eg}, 1998S \citep{fassia2001}, 
and 2007rt \citep{trundle07rt}, implying that the progenitors of each of 
these SNe may have been in a similar state shortly before core collapse.

The dense CSM and large luminosities associated with many SNe~IIn  
have led a number of authors to suggest that at least some SNe~IIn are 
associated with progenitors that experienced LBV-like mass loss shortly 
before core collapse (see e.g., \citealt{chu99-78K, salamanca2000-97ab, 
chugai95g, chugai94w}). This possible connection was considerably 
strengthened following the direct identification of the progenitor of 
the normal Type IIn SN 2005gl on archival {\it Hubble Space Telescope} 
images \citep{gal-yam2007, gal-yam09}. While the diagnostics above are 
not ubiquitous for all SNe~IIn\footnote{Note that the low-resolution 
spectrographs typically employed to observe SNe lack sufficient resolution 
to resolve narrow (FWHM $\la$ 150 \kms) absorption lines.}, many of the 
signatures (high-density CSM, episodic mass loss with \mdot\ $\approx 10^{-2}$ 
\msun\ yr$^{-1}$, and $\sim$100 \kms\ progenitor wind speed) are shared with 
\sn, suggesting that it too had an LBV-like progenitor.

As mentioned above, an alternative way to generate a wind-density parameter 
that is increasing with distance from the progenitor is to increase 
the progenitor wind speed in the years leading up to the SN event. A 
relatively mild change in the wind speed by a factor of $\sim$3 could, 
in fact, explain the observed increase in luminosity. During the 
post-main-sequence evolution of very massive stars a slightly more extreme 
transition is expected as stars exit the RSG or LBV phase, with wind speeds 
of $\sim$10$-$100 \kms, to become Wolf-Rayet stars, with typical 
wind speeds of $\ga$ 1000 \kms.
As this fast wind expands into the slower wind it will 
sweep up a thin, dense shell at the interface of the two winds and thus 
create a wind-blown bubble (e.g., \citealt{dwarkadas2005}). A SN 
exploding in one of these bubbles will appear as a normal SN until the 
ejecta reach the edge of the bubble. As the ejecta overtake the thin shell 
the appearance of the SN will dramatically change as CSM interaction 
begins to provide a significant contribution to the broad-band luminosity. 
This is the precise scenario proposed for the Type IIn SN~1996cr, for which 
\citet{bauer08-96cr} found an optical luminosity that is clearly decreasing 
between 1996 and 1999, but the radio emission sharply increases in late 
1997 while X-ray emission abruptly turns on some time 
between 1998 Mar.\ and 2000 Jan. The X-ray emission is especially 
interesting because after the abrupt initial rise, it continues to 
rise for at least the next seven years, a feature which had not previously 
been observed for any X-ray SN. A similar scenario is proposed for 
SN~2001em, classified as a Type Ic, but 
which showed unexpectedly strong radio and X-ray emission $> 2$ yr after 
explosion, as well as a FWHM $\approx$ 1800 \kms\ H$\alpha$ emission 
line \citep{soderberg04}. Each of these late-time peculiarities can be 
understood if the progenitor of SN~2001em lost the remainder of its 
hydrogen envelope during a phase of large mass-loss $\sim$(2--10) 
$\times 10^{-3}$ \msun\ yr$^{-1}$, about (1--2) $\times 10^{3}$ yr prior to 
core collapse \citep{chugai+chevalier}. After the hydrogen envelope has been 
lost, a fast wind from the progenitor sweeps the hydrogen into a shell. 
The interaction of the Type Ic ejecta with this swept-up shell give rise 
to the unusual late-time emission.

Can the long optical rise be the result of \sn\ exploding inside a 
wind-blown bubble? There is a somewhat sharp rise in the optical 
($>$ 2 mag brightening during the 68 days between the last DS 
non-detection and the first CRTS detection of the SN) followed by a 
slow and steady increase over the next $\sim$400 days, which may hint 
that the SN exploded in such a wind-driven cavity. However, we do not favour 
this model. In the wind-blown bubble scenario, after peak the thin shell 
has been overtaken by the ejecta, and the luminosity is powered by the 
interaction of the ejecta with the slower, older wind. To explain this, 
the wind on day 711 requires \mdot\ $\approx$ 10$^{-2}$ \msun\ yr$^{-1}$ and 
$v_w \approx 100$ \kms, which requires LBV-like mass loss (see above). 
The wind-blown bubble scenario therefore seems to unnecessarily complicate 
the scenario discussed above, which is needed to explain the late-time 
observations, as it requires the transition from a LBV to a Wolf-Rayet 
star. Yet the addition of this transition does not clearly model any 
observed properties that could not otherwise be explained. 

Additionally, we see no evidence for a normal SN in the years leading up to 
the initial detection of \sn\ as would be expected in the wind-blown 
bubble scenario. While it would be impossible to completely rule out a SN 
prior to the first detection of \sn\ on 2007 Sep.\ 13 (for instance, the 
field is virtually unobservable for $\sim$3 months each year while it is 
behind the Sun), DS images dating to 2004 show no evidence for such an 
event. These DS images have similar depths to those quoted in Section~2.1, 
which at the distance of the host galaxy corresponds to an absolute magnitude 
of $M_i \approx -15.4$ mag. This detection threshold is well below the mean 
peak magnitude for stripped-envelope SNe: $M_R = -17.01, -16.38,$ and 
$-17.04$ mag for SNe~Ib, Ic, and IIb, respectively \citep{wli-rates-LF}. 
Stronger constraints on the presence of a  stripped-envelope SN prior to 
2007 Sep.\ 13 are based on the individual DS images. There are DS images of the 
field of \sn\ from 2004 Apr.\ and Jul., 2005 May, Jun., and Jul., 2006 
May and Jun., and 2007 Apr., May, Jun., and Jul. Adopting the mean peak 
magnitudes and mean light curves for stripped-envelope SNe from 
\citet{wli-rates-LF}, we find that on average (note that there is a great deal 
of dispersion in the peak absolute magnitude and duration of stripped-envelope 
SNe), SNe~Ib and IIb are observable for $\sim$70 days above the 
DS detection threshold, while SNe~Ic are observable for only $\sim$45 
days above the DS detection threshold. In addition to not being 
detected during any of the months listed above, we can also rule out SNe Ib 
and IIb in the $\sim$2 months prior to those listed above, while SNe Ic can 
be ruled out in the $\sim$1.5 months prior to those above. Therefore, in 
the $\sim$3.5 years prior to \sn\ we can rule out SNe~Ib or IIb in 
$\sim$21 out of 42 months, while SNe~Ic can be ruled out for $\sim$18 
out of 42 months. 

Perhaps the most intriguing points about the progenitor of \sn\ are the 
implications for the progenitors of the very long-lived SNe~IIn 
discussed in Section 4.2, such as SN~1988Z. It is interesting that 
episodic, LBV-like eruptions, which we argue occurred shortly before \sn, may 
be applicable to SNe~1988Z, 1995N, and 1986J, each of which has relatively 
poor constraints on the actual date of explosion. 

SN~1988Z was first observed after peak 
\citep{stathakis1991}, with an observational constraint 
on the rise time of $<$ 250 days \citep{turatto1993}. It was argued by 
\citet{turatto1993} that the rise time of SN~1988Z must have been 
short, in part because at the time no other known SN looked quite like 
SN~1988Z. \sn, which is a virtual clone of SN~1988Z at late times, does 
provide an example of a SN with a long rise time and a peak absolute magnitude 
that is brighter than the discovery magnitude of SN~1988Z. This negates 
the necessity of a short rise to peak for SN~1988Z. We submit that it is 
possible that SN~1988Z had a long rise time, though we note that with no 
observations during the 250 days prior to discovery, there is no way to 
definitely prove this claim one way or another. A long rise to peak would, 
however, further strengthen the similarity of \sn\ and SN~1988Z. 

Similarly, SNe~1986J and 1995N have poor constraints on the rise time. 
SN~1986J was discovered in the radio, with constraints on the explosion date 
ranging from 1982 to 1984; constraints on the optical rise are even worse, 
owing in part to the large extinction in the host galaxy, $A_V \approx$ 2 mag 
(see \citealt{rupen86J}). Typically, $\sim$10 months prior to discovery 
is adopted as the explosion date for SN~1995N (see \citealt{fox2000}). 
However, this age of 10 months is based on the H$\alpha$ profile of 
SN~1995N, which looked similar to the H$\alpha$ profile of SN~1993N 10 months 
after SN~1993N had been discovered \citep{benetti95N}. We note that SNe~IIn 
are a very heterogeneous subclass, and in our experience spectral ages of 
SNe~IIn are not reliable, especially at times greater than  
a few months. Thus, poor optical constraints mean that SNe~1986J, 1988Z, and 
1995N all may have had long rise times, by the standards of typical 
SNe~II whose rise times are $\la$ 1 week \citep{wli-rates-LF}.

\subsection{Similarities to Very Luminous SNe~IIn}

The long rise time and broad, symmetric evolution of \sn\ around peak 
optical emission is reminiscent of the Type IIn VLSNe 2006gy 
and 2008fz \citep{ofek06gy, smith07-2006gy, drake09-2008fz}. Applying the 
``shell shock'' model of \citet{smith+mccray07}, which was originally 
developed to explain the light curve of SN 2006gy, does not provide a 
physically 
plausible scenario for \sn\ because the relatively low \mdot\ and large 
radius at peak imply that the CSM shell was not opaque. The ``shell shock'' 
model argues that the extreme luminosity of these SNe is powered via a 
shock running into a dense, optically thick shell. Radiation from the 
shock is thermalised by the optically thick gas and must diffuse out 
of the shell. 

Nevertheless, \sn\ does share a few similarities with the 
VLSNe~IIn. Like SNe~2006gy and 2006tf, \sn\ shows evidence for a 
$\sim$100 \kms\ pre-shock CSM wind (see \citealt{smith07-2006gy,smith06tf}), 
and like SNe~2006gy, 2006tf, and 2008fz, 
\sn\ experienced a period of enhanced mass loss during the decades 
prior to explosion. The difference between these systems is that the 
mass-loss rate from \sn, \mdot\ $\approx 0.01$ \msun\ yr$^{-1}$, was less extreme 
than that from SNe 2006gy, 2006tf, 2008fz, where \mdot\ $\approx 1.0$ 
\msun\ yr$^{-1}$ (e.g., \citealt{smith09-2006gy}). Also, the time between 
the period of enhanced mass loss 
and the SN was longer for SN~2008iy, $\sim$ 55 years, than for SNe~2006gy and 
2006tf, $\la 10$ years \citep{smith09-2006gy, smith06tf}. The result of these 
differences is that \sn\ did not reach as extreme a peak luminosity and 
took considerably more time to reach peak optical output than the 
VLSNe~IIn. Finally, if the late-time NIR excess is the result 
of an IR dust echo, this would be similar to SN~2006gy \citep{smith08-2006gy, 
agnoletto09, miller06gy}. \citet{smith08-2006gy} and \citet{miller06gy} 
argue that a giant eruption $\ga$1000 year prior to SN~2006gy could 
potentially create 
the dust shell giving rise to the echoes, and a similar shell around 
\sn\ may explain the observed NIR excess.

The connection between VLSNe and very massive stars, such as $\eta$ Car, was 
first made by \citet{smith07-2006gy} because the large mass-loss rates 
needed to explain the extreme optical luminosity are reminiscent 
of $\eta$ Car during the great eruption of 1843. It is interesting to note 
that were $\eta$ Car to explode as a SN today, its appearance would be less 
like the VLSNe IIn discussed above and more similar to \sn. The 
reason for this is that with the eruption happening $>$150 yr ago, 
the ejected shell surrounding $\eta$ Car has had sufficient time to 
expand ($R \approx {\rm few} \times 10^{17}$ cm) and become clumpy 
\citep{morse1998}, with an optical depth that is now 
near unity \citep{davidson2001}. Thus, $\eta$ Car does not have the compact 
($R \la {\rm few} \times 10^{15}$ cm), very optically thick shell 
required to generate an extreme peak luminosity following a SN. Instead, 
as the SN ejecta sweep through the large shell,\footnote{The shell around $\eta$ Car 
is much bigger than the one around \sn\ because the wind speeds are 
greater, $v_{w, \eta Car} = 600$ km s$^{-1}$ \citep{smith06-etacarH2}, and the 
nebula is older, $\sim$165 yr relative to $\sim$55 yr.} they will 
successively overtake more and more mass within the clumpy CSM. This will 
likely give rise to a long rise time, similar to what was seen in SN~2008iy.

\subsection{The Host of \sn}

At the adopted distance to \sn, our detection of the host galaxy in stacked 
SDSS images (see Section~2.3) means that the host has an 
absolute magnitude of $M_r \approx -13.7$ mag. This is significantly 
fainter than the Small Magellanic Cloud (with $M_V = -16.9$ mag). We do not 
detect narrow emission lines from the host in our SN spectra in order to 
make a direct measurement of the metallicity. The prospects for such a 
measurement in the near future are not good: if \sn\ continues to evolve 
like SN~1988Z, light from the SN may dominate over light 
from the host galaxy for years to come. 

We can, however, estimate the 
metallicity based on the absolute magnitude of the host. \citet{lee2006} 
use 27 dwarf irregular galaxies to determine the luminosity-metallicity 
relation for low-mass galaxies, and following 
from their Equation~1, we find $12 + \log({\rm O/H}) 
\approx 7.7$, assuming $M_B \approx M_r$. The uncertainty in this value 
is large, perhaps as great as $\pm$ 0.3 dex, based both on our assumption 
of $B-r = 0$ mag for the host and the substantial scatter in the 
luminosity-metallicity relationship for faint galaxies (\citealt{lee2006} 
find a scatter of 0.16 dex), yet this nevertheless shows that 
\sn\ occurred in a metal-poor galaxy.

It has recently been suggested that the unusual SNe discovered by the 
non-targeted transient surveys, such as CRTS, preferentially occur in 
subluminous, possibly metal-poor host galaxies \citep{miller08es, 
drake09-2008fz}. \sn\ is yet another example of an unusual SN in a 
low-luminosity host. We caution, however, that a number of biases may be 
skewing initial impressions about the hosts of these unusual SNe. 
Preliminary calculations show that these VLSNe are rare 
\citep{miller08es,quimby2009}. Both the large peak luminosity and slow 
evolution of \sn\ mean that similar SNe would be easily detectable in 
the galaxies of targeted SN surveys. The lack of 
other examples of SNe with $\sim$400 day rise times suggests that \sn\ is 
also rare. The CRTS, however, does not employ image subtraction during 
their search for transients; instead they use aperture photometry to find new 
sources, or sources with large increases in flux, and flag those as possible 
transients. Consequently, their survey is biased 
toward the discovery of intrinsically bright SNe in faint host galaxies. 
From the Lick Observatory Supernova Search (LOSS), we know that SNe~IIn 
are rare, regardless of whether they are very luminous or have long
rise times \citep{wli-rates-rates}. Yet, 
even LOSS, which targets specific galaxies, may be biased in 
that it observes relatively few subluminous galaxies. New and upcoming 
surveys, such as the Palomar Transient Factory \citep{law09-ptf,rau09-ptf}, 
which employ image subtraction and survey large, non-targeted fields 
of view, will not suffer from the same biases discussed above, and hence 
will be in a better position to address whether these unusual 
SNe preferentially occur in subluminous host galaxies.

\section{Conclusions}

We have reported on observations of the Type IIn \sn, which took 
$\sim$400 days to reach peak optical output. There are few known SNe with 
optical rise times $\ga$ 50 days, and \sn\ is the first with a rise time 
$>$ 1 year. We argue that this long rise to peak is caused by the interaction 
of the SN ejecta with a dense CSM; radioactivity is unlikely to drive a 
400-day rise. Furthermore, 
the late-time optical decay, $\sim$0.4 mag (100 days)$^{-1}$, is slower 
than that of $^{56}$Co, which provides further evidence that radioactive 
heating is not a dominant energy source for \sn. Spectroscopically, 
\sn\ is very similar to SN~1988Z at late times. SN~1988Z is understood to 
have exploded in a dense, clumpy CSM \citep{chugai1994}, which we 
argue was also the case for \sn. We detect \sn\ in 
X-rays with a total luminosity $L_{\rm X} = (3.7 \pm 1.2) \times 10^{41}$ 
erg s$^{-1}$, which is similar to that of the Type IIn SNe~1988Z and 1995N 
\citep{fox2000}. Similar to other SNe~IIn, \sn\ has a growing NIR-excess 
at late times. \sn\ had a peak absolute magnitude of $M_r \approx -19.1$ 
mag and a total radiated energy of $\sim 2 \times 10^{50}$ erg in 
the optical (assuming no bolometric correction). 

The steady increase in optical luminosity over a $\sim$400-day period 
means that the wind-density parameter, $w$, increased over a distance 
of $\sim$1.7 $\times 10^{16}$ cm from the SN. We propose two possible 
scenarios to explain this increase in $w$: (i) the progenitor 
experienced an episode of LBV-like, eruptive mass-loss $\sim$55 years 
prior to the SN, or (ii) the wind speed of the progenitor was 
increasing during the years leading up to core collapse. We prefer the 
former scenario, as the later adds unnecessarily complicated pieces to 
the puzzle without providing a unique solution. Our favoured scenario 
provides yet 
another piece of evidence that some SNe~IIn are connected to LBV-like 
progenitors (see \citealt{gal-yam2007}, and references therein). We 
find that the host of \sn\ is a subluminous dwarf galaxy, though we 
caution against premature conclusions that unusual SNe, specifically 
those that are very luminous or have very long rise times, 
preferentially occur in low-mass dwarf galaxies.

Finally, we close with some predictions for the late-time behaviour of 
\sn. There are 
examples of SNe~IIn whose luminosity dramatically drops after the 
SN ejecta overtake the dense CSM (e.g., SN~1994W; \citealt{sollerman94w}). 
However, the similarities to SN~1988Z suggest that \sn\ could continue 
interacting, and thus remain luminous, for several years. This would allow 
long-term monitoring in the radio, X-ray, and optical, like SN~1988Z 
\citep{williams2002, schlegel2006, aretxaga88z}. We predict that \sn\ 
is a luminous radio source, like SN~1988Z \citep{vandyk88z}, though we note 
that at $z = 0.0411$ deep observations may be necessary for detection. The 
increasing $K_s$-band luminosity from day 560 to 710 
is likely due to the presence of dust, and we predict that 
\sn\ will also be luminous in the mid-IR ($\sim$3$-$5 $\mu$m). The data are 
insufficient to distinguish between newly formed dust or dust that was 
present prior to the SN, but future medium- to high-resolution 
spectroscopy could distinguish between these two cases. If new dust is 
being formed, it should result in a systematic blueshift in the line 
profiles, as the dust creates an optically thick barrier to radiation 
from the receding SN ejecta.


\bigskip

A.A.M. would like to thank D. Poznanski for useful discussions that 
helped improve this paper and M. Modjaz for discussions concerning the 
metallicity of the host.
We thank Neil Gehrels for approving the {\it Swift} ToO request for 
\sn, which was submitted by Dave Pooley, and the {\it Swift} team for 
scheduling and obtaining those observations.
We acknowledge the use of public data from the {\it Swift} archive.
Cullen Blake, Dan Starr, and Emilo Falco assisted with the 
operation of PAIRITEL. 
We wish to thank the following Nickel observers: P. Thrasher, M. Kislak, 
J. Rex, J. Choi, I. Kleiser, J. Kong, M. Kandrashoff, and A. Morton. 
We thank the referee, Nikolai Chugai, for suggestions that 
helped to improve this paper.

A.A.M. is supported by the NSF Graduate Research Fellowship Program 
and DOE grant \#DE-FC02-06ER41453. 
J.S.B.'s group is partially supported by NASA/{\it Swift} grant
\#NNX08AN84G.   A.V.F.'s group is grateful for the financial
assistance of NSF grant AST--0908886 and
the TABASGO Foundation.  
PAIRITEL is operated by the Smithsonian Astrophysical
Observatory (SAO) and was made possible by a grant from the Harvard
University Milton Fund, the camera loan from the University of
Virginia, and the continued support of the SAO and UC Berkeley. The
PAIRITEL project is partially supported by NASA/{\it Swift} Guest
Investigator Grant \#NNX08AN84G.
This publication makes use of data products from the Two Micron All
Sky Survey, which is a joint project of the University of
Massachusetts and the Infrared Processing and Analysis
Center/California Institute of Technology, funded by the National
Aeronautics and Space Administration (NASA) and the National Science
Foundation (NSF).
Some of the data presented herein were
obtained at the W. M. Keck Observatory, which is operated as a
scientific partnership among the California Institute of Technology,
the University of California, and NASA; it was made possible by the generous
financial support of the W. M. Keck Foundation. The authors wish to
recognise and acknowledge the very significant cultural role and
reverence that the summit of Mauna Kea has always had within the
indigenous Hawaiian community; we are most fortunate to have the
opportunity to conduct observations from this mountain.



\label{lastpage}

\end{document}